\pdfoutput=1
\documentclass[%
reprint,
showpacs,
amsmath,amssymb,
aps,
prb,
]{revtex4-1}

\usepackage{graphicx}
\usepackage{bm}
\usepackage{braket}
\usepackage[caption=false]{subfig}

\DeclareMathOperator{\tr}{tr}

\begin{document}
\title{Hybridization of Higgs modes in a bond-density-wave state in cuprates}

\author{Zachary M. Raines}
\author{Valentin G. Stanev}
\affiliation{
Condensed Matter Theory Center, Department of Physics, University of Maryland, College Park, Maryland 20742-4111, USA}
\author{Victor M. Galitski}
\affiliation{Joint Quantum Institute and Condensed Matter Theory Center, Department of Physics, University of Maryland, College Park, Maryland 20742-4111, USA}
\affiliation{School of Physics, Monash University, Melbourne, Victoria 3800, Australia}

\begin{abstract}
Recently, several groups have reported observations of collective modes of the charge order present in underdoped cuprates. Motivated by these experiments, we study theoretically the oscillations of the order parameters, both in the case of pure charge order, and for charge order coexisting with superconductivity. Using a hot-spot approximation we find in the coexistence regime two Higgs modes arising from hybridization of the amplitude oscillations of the different order parameters. One of them has a minimum frequency that is within the single particle energy gap and which is a non-monotonic function of temperature.  The other -- high-frequency -- mode is smoothly connected to the Higgs mode in the single-order-parameter region, but quickly becomes overdamped in the case of coexistence. We explore an unusual low-energy damping channel for the collective modes, which relies on the band reconstruction caused by the coexistence of the two orders.  For completeness, we also consider the damping of the collective modes originating from the nodal quasiparticles. At the end we discuss some experimental consequences of our results.
\end{abstract}


\maketitle

\section{Introduction}
Despite the decades of intense research efforts, superconductivity in cuprates remains a profound mystery. However, recently there has been a lot of progress in clarifying and refining the phase diagram of these materials experimentally.~\cite{Taillefer2010a,Keimer2014} In particular, there is growing evidence of a charge order existing in the pseudogap state of several cuprate families,\cite{Ghiringhelli2012, Chang2012, Achkar2012, Coslovich2013,  Fujita2014, Comin2014, Forst2014} which also coexists and competes with superconductivity at lower temperatures. Furthermore, it appears that this order has a non-trivial $d$-wave phase factor,\cite{Fujita2014,Comin2014} implying that within a one-band model of copper sites it describes ordering entirely on the links. For this reason it has been dubbed ``bond-density wave" (BDW). 
Recently, several groups employed time-domain reflectivity\cite{Demsar1999} as a tool to study this order,\cite{HintonPRL,Hinton2013,Torchinsky2013} and, in particular, its collective modes. In some cases they were able to extract the amplitude and phase oscillations and to track them as the system became superconducting.
These results can provide valuable insights into the physics of both pseudogap and superconducting states, and, thus, it is desirable to have a better theoretical understanding of the possible collective modes of these systems. One particularly interesting point is that the coexistence of charge order and superconductivity makes possible the direct observation of the superconducting Higgs mode, as first pointed out in the pioneering work of Littlewood and Varma.~\cite{Littlewood1981}

In this work we present a theoretical study of the Higgs modes, or oscillations of the amplitude, of the order parameters in underdoped cuprates.  We consider both the pure BDW state, as well as the coexistent BDW-superconductivity phase. We use the so-called ``hot-spot'' model\cite{Metlitski2010, Sachdev2013,Efetov2013,Sau2013, Allais2014,Allais2014a,Thomson2014, Raines2015a} of the pseudogap phase, which is based on a picture of a metallic state close to a magnetic instability, and considers the physics of the special points on the Fermi surface connected by the magnetic ordering vector. Although relatively simple, this model has seen extensive use recently, as it naturally leads to coexistence between BDW and superconductivity, and also correctly predicts the $d$-wave phase factor of the charge order.~\cite{Sachdev2013, Efetov2013,Sau2013} 

Our results provide a general framework for identifying and understanding order-parameter collective modes of the system. In the single-order phase (i.e., only BDW or superconductivity) we find, as expected, a single amplitude mode, which is coupled with the quasiparticle continuum and is always damped. However, the coexistence regime is much more interesting -- the fluctuations of the different order parameters become intertwined.~\cite{Littlewood1981} 
As a consequence, in this region we find two Higgs modes, which represent coupled oscillations of the order parameters. One of the modes is slow, with frequency well below the amplitude of the order parameters, but which is, nevertheless, weakly damped. The other mode is pushed inside the high-energy continuum, and quickly becomes overdamped. We follow the slow mode in the entire coexistence phase, and find its frequency to be a non-monotonic function of temperature.
This mode is weakly damped through an unusual low-energy decay channel for the antinodal qausiparticles, caused by the coexistence of the two orders and the associated band reconstruction. Even more unusually, this damping initially increases with the decrease of temperature.
To account for the damping from the gapless degrees of freedom present at the nodal regions we develop a phenomenological time-dependent Ginzburg-Landau theory. We demonstrate that, by allowing for significant damping, the in-gap mode is strongly suppressed, while the frequency of the high-energy mode is brought down.
Our results provide a characterization of the amplitude modes of the coexistent superconductivity-BDW system which can be compared with the experimental data and used to identify the appropriate Higgs modes of the system.

\section{Microscopic calculation of collective modes frequencies}



We will consider here the collective modes in a 2D ``hot-spot" model.
Such a model can be obtained as a low-energy theory from the 2D t-J-V model,\cite{Metlitski2010,Sachdev2013,Efetov2013,Sau2013} which contains hoppings $t_{(1/2/3)}$ on a square lattice as well as nearest-neighbor exchange and Coulomb interactions $J$ and $V$.
Specifically, one projects the lattice theory onto regions in the vicinity of 8 ``hot spots'' where the Fermi surface intersects the magnetic Brillouin zone boundary.
In the vicinity of these hot spots the nearest-neighbor interactions $J$ and $V$ can be approximated by constants.
Time reversal symmetry allows the problem to be reduced to considering fermions near 4 inequivalent hot spots where, in the channels of interest, the interactions take the form
\begin{gather}
    \mathcal{H}^\Delta_\text{int} = \frac{g_s}{4} \sum_{k,p,q} \Psi^\dagger_{k+q,a} \check V_\Delta \Psi_{k, a} \Psi^\dagger_{p-q,b} \check V_\Delta \Psi_{p,b},\\
    \mathcal{H}^\phi_\text{int} = \frac{g_c}{4} \sum_{k,p,q} \Psi^\dagger_{k+q,a} \check V_\phi \Psi_{k, a} \Psi^\dagger_{p-q,b} \check V_\phi \Psi_{p,b},
\end{gather}
where $\Psi_{a,b}$ are Nambu spinors in pairs of hot-regions separated by the antiferromagnetic wave-vector $\vec K = (\pi, \pi)$ and $g_s$ and $g_c$ are the non-retarded components of the interaction in the superconducting and bond-density wave (BDW) channels, respectively. $\check{V}_\Delta$ and $\check{V}_\phi$ are the vertices for pairing in the superconducting and bond-density-wave channels.
Their explicit forms for the system studied here are shown in Eq.~\ref{eq:modeldef}.

Due to the $d$-wave symmetry of the order parameters one can further restrict attention to $2$ of the $8$ hot regions.~\cite{Sau2013}
The interaction terms can be decoupled via a Hubbard-Stratonovich transformation.
In the usual manner, the saddle point of the zero-frequency terms of the decoupling fields leads to a mean-field theory, which in this case has mean-field Hamiltonian
\begin{equation}
    \mathcal{H} = \sum_{\vec k} \Psi^\dagger_{\vec k} \check H_{\text{MF}}(\vec k) \Psi_{\vec k} + \frac{2}{g_s} |\Delta|^2 + \frac{2}{g_c} |\phi|^2,
    \label{eq:H0}
\end{equation}
where now $\Psi$ is a Nambu spinor $(c_{k1\uparrow}, c_{k2,\downarrow}, c_{-k2\downarrow}^\dagger, c_{-k1\uparrow}^\dagger)^T$ describing one pair of hot spots.
The mean-field Hamiltonian describes two species (denoted $1$ and $2$) of spinful fermions which pair only with each other.
Specifically,
\begin{equation}
\begin{gathered}
    \check{H}_{\text{MF}} = \check{H}_0 + \Delta \check{V}_\Delta + \phi \check{V}_\phi,\\
    \check{H}_0 =
    \mathrm{diag}(\xi_1, \xi_2) \otimes \hat \tau_z,\\
    \check{V}_\Delta = \hat \rho_1 \otimes \hat \tau_3,\quad
    \check{V}_\phi = \hat \rho_0 \otimes \hat \tau_1,
\end{gathered}
\label{eq:modeldef}
\end{equation}
where $\hat\tau_i$ and $\hat\rho_i$ are Pauli matrices acting in particle-hole space and species space, respectively, and $\Delta$ describes $d$-wave superconductivity while $\phi$ is the BDW order.~\cite{Sau2013}
Here, and in what follows, $\check{M}$ denotes a matrix in the $4\times4$ Nambu-hot-spot space, and $\hat M$ a $2\times 2$ matrix.
The self-consistency equations associated with Eq.~\ref{eq:modeldef} are
\begin{equation}
\begin{gathered}
    \Delta = \frac{g_s}{4} T \sum_k \tr \check{V}_\Delta \check{G}_k,\\
    \phi = \frac{g_c}{4} T \sum_k \tr \check{V}_\phi \check{G}_k,
    \label{eq:gap}
\end{gathered}
\end{equation}
where $\check{G}_k$ is the matrix Matsubara Green's function of the Hamiltonian in Eq.~\ref{eq:H0}, and $k=(i\epsilon_n, \vec k)$, with $\epsilon_n$ being a fermionic Matsubara frequency.
Here we have considered $\Delta$ and $\phi$ to be real and non-negative (they can always be brought to this form via a gauge transformation).

In the case of a hot-spot model of cuprates, the two species correspond to fermions within a vicinity of in-equivalent ``hot spots" in the Brillouin zone.
Close to the hot-spot points the electron dispersion can be modeled as $\xi_1(\vec k) = \xi_2(-\vec k) = v_f k_x + \gamma k_y^2$, where we include the curvature $\gamma$ as it plays an important role in breaking the degeneracy between the two orders and allowing coexistence.~\cite{Sau2013,Moor2014}

\begin{figure}
\centering
\includegraphics[width=0.8\linewidth]{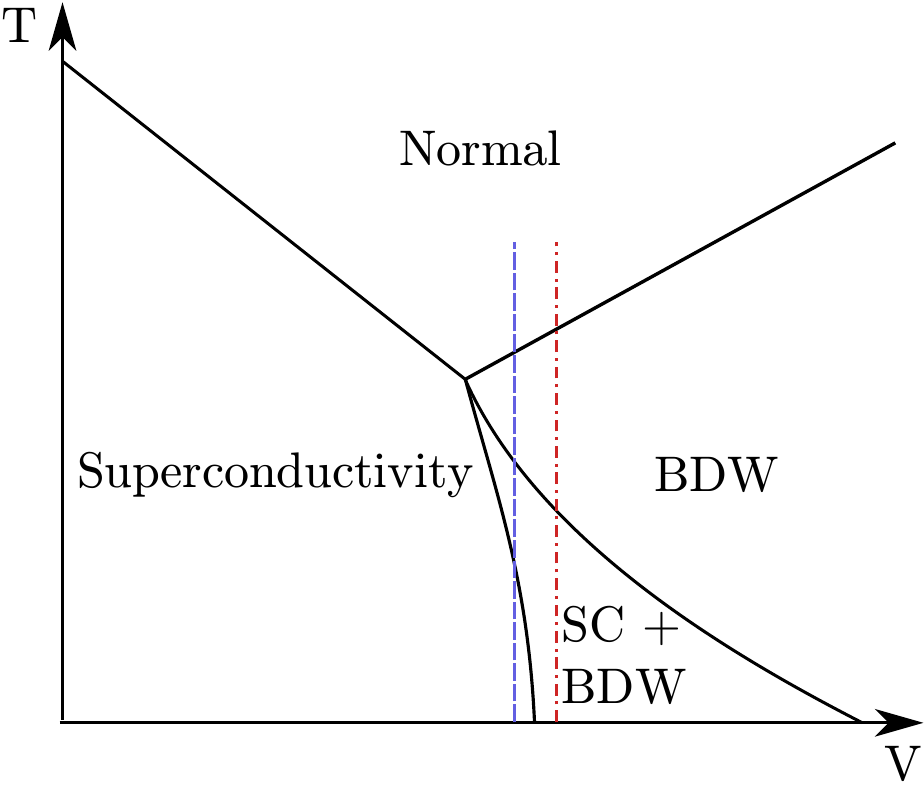}
    \caption{(Color online) Schematic phase diagram of the hot-spot model,~\cite{Sau2013, Raines2015a} which illustrates the transition from superconductivity to charge order, tuned by $V$ (nearest-neighbor Coulomb interaction).
In this work we consider the transition from BDW to BDW-superconducting mixed state along the ``trajectories" indicated by the dashed lines.
Depending on the exact value of $V$ the $T \rightarrow 0$ limit of the system could be either in a pure superconducting state (indicated by the blue dashed line), or in a mixed state (red dot-dashed line).
\label{fig:phase}}
\end{figure}

We follow Ref.~\onlinecite{Sau2013} by choosing units where $v_f=1$, $\gamma=1/\Lambda=\pi$, with $\Lambda$ being the hot spot cutoff, and parametrize $\{g_c, g_s\} = 3J \pm 4V$ with $J=1.2$.
Note that $V$ strengthens the interaction in the charge channel, while decreasing the interaction in the superconducting channel, and thus can be used to tune the coexistence (as depicted in Fig.~\ref{fig:phase}).
We consider two qualitatively different cases of coexisting charge order and superconductivity as depicted in Fig.~\ref{fig:phase}: one where charge order disappears for some finite temperature below the superconducting $T_c$ ($V = 0.2$), and one where charge order survives all the way down to $T=0$ ($V = 0.21$).
In both of these cases, the BDW order will onset at a temperature $T_\text{BDW} > T_c$.
The competition between the two orders can be readily confirmed by a decrease in $\phi$ below the superconducting $T_c$.


The ordering vector of the BDW is determined by the separation in the Brillouin zone of the hot spots being paired.~\cite{Metlitski2010,Sau2013,Sachdev2013,Allais2014,Allais2014a,Thomson2014} It is important to note that we are considering a BDW with ordering vector $(Q, Q)$, which is known to be the leading instability of this simple model.~\cite{Sau2013,Sachdev2013,Allais2014,Allais2014a,Thomson2014}
This is different from the experimentally observed bond-oriented ordering directions $(Q, 0)$ and $(0, Q)$, which correspond to a different choice of hot spots for the BDW pairing to occur between. It is possible to stabilize the $(Q, 0)$ and $(0, Q)$ orders,~\cite{Allais2014, Chowdhury2014, Thomson2014} but at the price of significantly complicating the model, and we will not pursue these modifications here.
We expect that most of our results and conclusions are applicable to the $(Q, 0)$/$(0, Q)$ orders as well.

\subsection{Hybridized Higgs modes}
\label{sec:energygap}
The collective modes of coexisting charge-density-wave and superconducting states have been studied theoretically previously,~\cite{Littlewood1981,Littlewood1982,Browne1983, Lei1, Lei2, Tutto, Cea} and  we apply the methods developed in these earlier works.
In general, the collective modes of the system are described by a $5\times 5$ matrix, which includes the amplitude and the phase modes of each order parameter, as well as the density oscillations of the fermions.
However, this matrix factorizes into two decoupled sectors, \cite{Browne1983} with a $2 \times 2$ block describing the interacting amplitude modes, and the other -- $3 \times 3$ -- block describing the order parameters phases coupled to each other, as well as to the fermionic density.\footnote{The oscillations of the phase of superconductor are usually pushed up to plasma frequencies by coupling with the Coulomb interaction. In contrast, the phase mode of an incommensurate charge order is theoretically a Goldstone mode of the system, but in real materials this degree of freedom is usually pinned by disorder.}
This being the case, we devote our attention to the amplitude mode sector.
In particular, we consider amplitude fluctuations of these order parameters with finite frequency $\omega$, but zero wave-vector.
Doing so allows us to calculate the mass of the collective modes: the minimum energy required to excite the collective modes of the ordered state.

Returning to the Hubbard-Stratonovich decoupling of the hot-spot model's interactions, inclusion of the finite-frequency components of the decoupling fields leads to the action
\begin{multline}
    S =
    S_\text{MF} +
    \sum_{k, \vec q, \omega_m} \bar\Psi_{\vec k + \vec q, \epsilon_n + \omega_m} \left(
    \Delta_{\vec q,\omega_m} \check{V}_\Delta + \phi_{\vec q, \omega_m} \check{V}_\phi \right)\Psi_{\vec k, \epsilon_n}\\
    + \frac{2}{g_s} \sum_{\vec q, \omega_m} |\Delta_{\vec q, \omega_m}|^2 + \frac{2}{g_c} \sum_{\omega_m, \vec q} |\phi_{\vec q, \omega_m}|^2,
\end{multline}
where $S_\text{MF}$ is the action corresponding to Eq.~\ref{eq:H0} and we are working in imaginary time.
We have kept here the fluctuations $\Delta(\tau)$, $\phi(\tau)$ which are along the direction of $\Delta,\phi$ in the complex plane,\footnote{$\Delta(\tau)$ and $\phi(\tau)$ may be treated, then, as real fields since they may always be brought to lie along the real axis via a gauge transformation.} corresponding to the amplitude modes.\footnote{This is a parametrization in terms of longitudinal and transverse modes such as considered in Refs.~\onlinecite{Littlewood1982,Browne1983} as opposed to radial and angular modes (c.f. D. Pekker and C.M. Varma, Annu. Rev. Condens. Matter Phys. 6, 269 (2015)).}

Particularly we will be interested in the $2\times 2$ matrix collective mode propagator
\begin{equation}
    D_{ij}(\omega_m, \vec q) = \langle O_{i,\omega_m, \vec q} O_{j, -\omega_m, - \vec q} \rangle,
    \label{eq:D}
\end{equation}
where $O_{1, \omega_m, \vec q} = \Delta_{\omega_m, \vec q}$ and $O_{2, \omega_m, \vec q} = \phi_{\omega_m, \vec q}$,
and the related object $D^R_{ij}(\omega, \vec q) = D_{ij}(i\omega_m \to \omega + i0^+, \vec q)$, which can be obtained via analytic continuation.
The off-diagonal elements of this matrix are in general non-zero and this is what leads to the hybridization of collective modes.
The poles of the retarded propagator $\hat D^R$ will describe the on-shell collective mode energies.

After integrating out the fermionic degrees of freedom, $\hat D(\omega_m, \vec q)$ can be expressed (at the quadratic level) as
\begin{equation}
    \hat{D}^{-1}(\omega_m, \vec q) = (\hat D^0)^{-1} - \hat Q(\omega_m, \vec q),
    \label{eq:Dinv}
\end{equation}
where we have defined
\begin{equation}
    \hat{D}^0 \equiv
    \frac{1}{4}
    \begin{bmatrix}
    g_s& 0\\
    0 & g_c
    \end{bmatrix}.
\end{equation}

Here $Q_{ij}$, also a $2\times 2$ matrix, is the self-energy of the collective modes due to the fermionic quasiparticles (this treatment is equivalent to obtaining the generalized susceptibilities of the order parameters within the RPA approximation).
Since, $\hat D^0$ is already known, $\hat Q$ is the object of interest.

Specifically, $\hat Q$ is given by
\begin{equation}
    Q_{ij}(i \omega_m, \vec q) = - T \sum_{\vec k, \epsilon_n}\tr \left[\check{G}(\vec k, \epsilon_n) \check{V}_i \check{G}(\vec k - \vec q, \epsilon_n - \omega_m) \check{V}_j\right],
    \label{eq:Q}
\end{equation}
where $i,j \in \{\Delta, \phi\}$.
After performing the fermionic Matsubara sums in Eq.~\ref{eq:Q} we analytically continue the bosonic frequency to the real axis, in order to obtain the finite-temperature, retarded self-energy $Q^R(\omega, \vec q)$.

The long-wavelength frequencies of the amplitude modes are given by the solutions of
\begin{equation}
\det[(\hat D^0)^{-1} - \hat Q^R(\omega_0 - i \Gamma_0, \vec q \to 0)] = 0,
\label{eq:detD}
\end{equation}
where
\begin{gather}
\omega_0 \equiv \mathrm{Re}[\omega(q \to 0)],\\
\Gamma_0 \equiv -\mathrm{Im}[\omega(q \to 0)],
\end{gather}
are, respectively, the mass and the decay rate of the Higgs mode in the long wavelength limit.
The in-gap collective modes, are those for which $\omega_0 < 2\min(\phi, \Delta)$.

One can explicitly show that the diagonal components of $\hat Q^R$ reproduce the usual $2\phi$/$2\Delta$ amplitude modes\cite{Littlewood1981} in the limit where one of the order parameters vanishes.
However, in our case, we focus our attention on the eigenmodes of the response function, which describe hybridized modes of the system\footnote{A similar framework was recently used in Ref.~\onlinecite{Cea}. However, the focus of that work was on the effects of the superconducting gap on the charge order, and the off-diagonal terms of $\hat Q^R$ (and thus the mixing) were assumed to be small.} and which cannot be obtained from purely considering the superconducting and BDW susceptibilities.

Because we are interested in weakly damped oscillations such that it makes sense to describe them as collective modes, we are able to employ a technique to determine the complex frequency of the oscillations from considerations of the response function on the real frequency line.
In particular, we obtain the real part of the frequency as the solution to the equation $\mathrm{Re}[\lambda(\omega_0)]=0$ where $\lambda$ is a solution to the eigenvalue problem
\begin{equation}
\left[(\hat D^0)^{-1} - \hat Q^R(\omega) - \lambda(\omega) \hat I\right]
\begin{pmatrix}
\Delta_\omega\\
\phi_\omega
\end{pmatrix}
= 0.
\label{eq:eigval}
\end{equation}
The imaginary part of the frequency can then be calculated by expanding the eigenvalue as a function of complex $\omega$ about the real frequency.~\cite{Kulik,Littlewood1982}
We defer analysis of the imaginary part (shown in Fig.~\ref{fig:damping}) until Sec.~\ref{sec:antinode} and focus now on the real part.

In order to track the temperature dependence of the collective modes, we explicitly solve the mean field equations for a range of temperatures and then calculate the collective mode frequencies at each temperature.
Below $T_\text{BDW}$, in the pure BDW phase, we find an amplitude mode starting at frequency $2\phi$, as expected.~\cite{Littlewood1981}
With the onset of superconductivity, another mode appears inside the gap. Physically, it represents coupled oscillations of the two order parameters, wherein pairs are excited in both the BDW and superconducting channels.
The mixing of the two orders arises due to the off-diagonal elements of $\hat Q^R$,  proportional to $\phi \Delta$.
Intuitively, one might anticipate the presence of such an in-gap mode by arguing that one could convert one type of pairing into the other at a smaller energy than it would take to completely break a pair.

\begin{figure}
\centering
    \includegraphics[width=\linewidth]{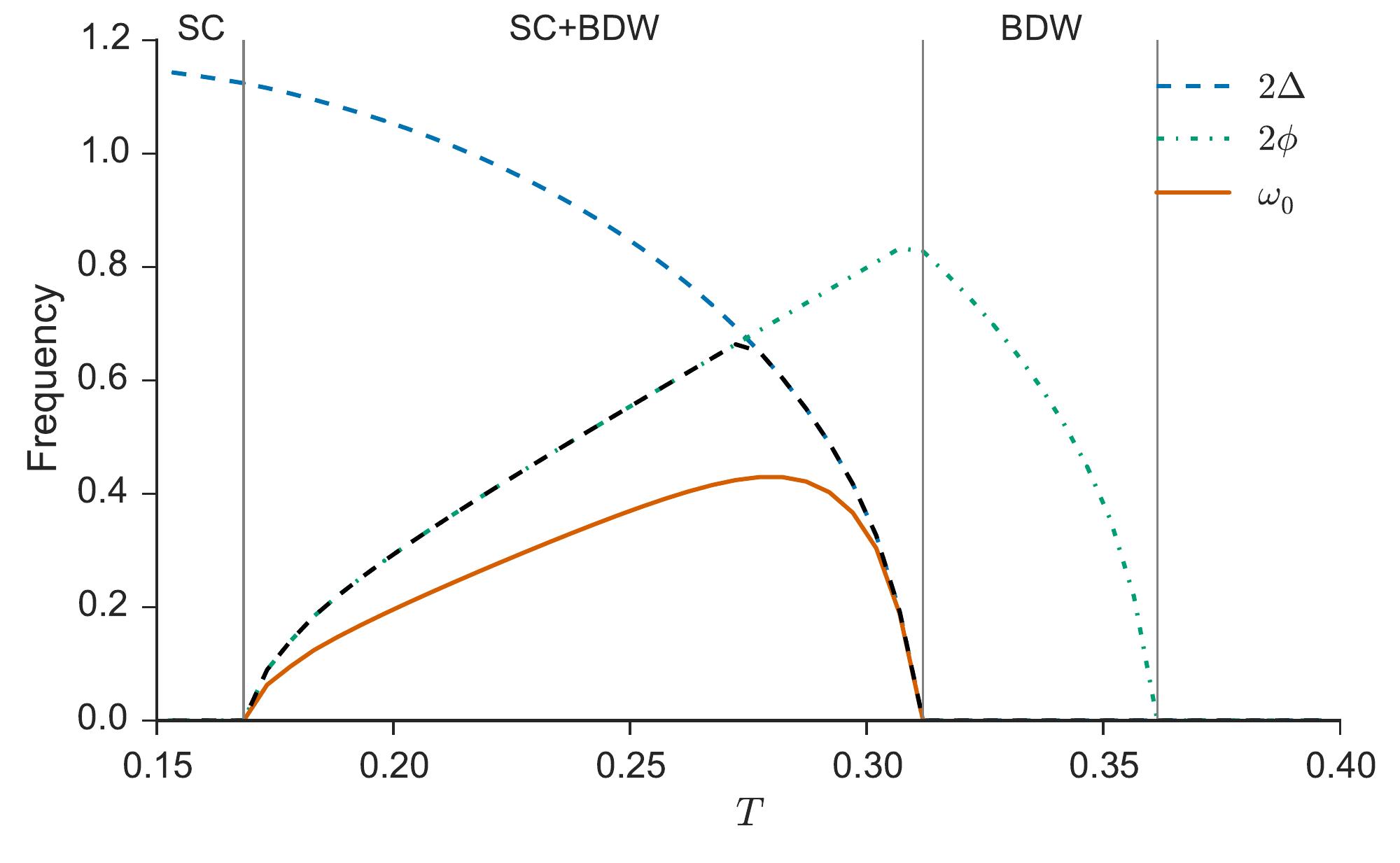}\\
    \includegraphics[width=\linewidth]{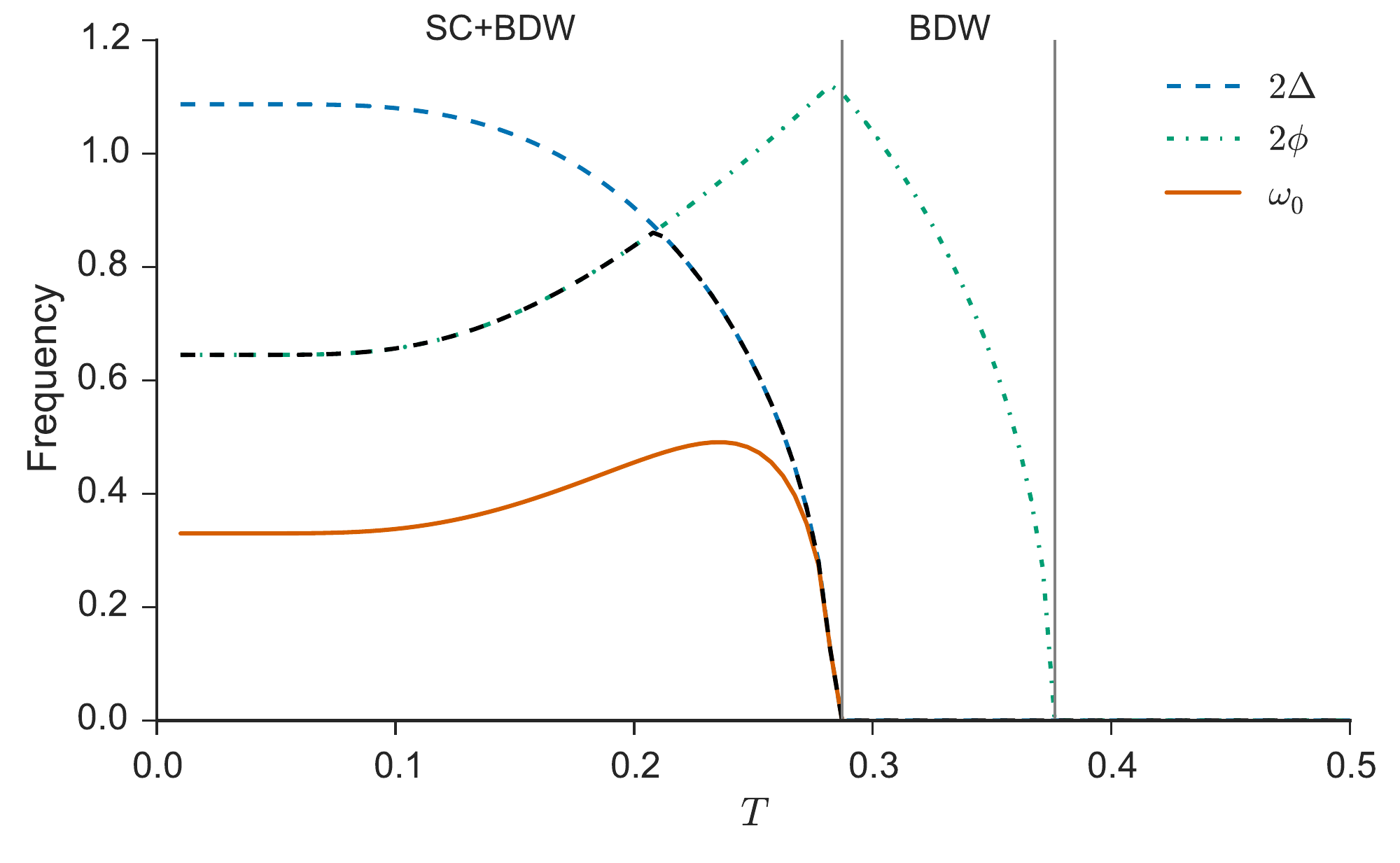}
\caption{(Color online) 
Mass of the in-gap hybrid Higgs mode $\omega_0 = \mathrm{Re}[\omega(q \to 0)]$ as obtained from Eq.~\ref{eq:detD}.
The frequency is plotted as a function of temperature for two different cases of $\phi(T \to 0)$ ($V = 0.2, 0.21$) as depicted by the dashed lines in Fig.~\ref{fig:phase}, using the units of Ref.~\onlinecite{Sau2013}.
A soft mixed mode emerges in both cases below the superconducting $T_c$.
For reference, twice the single-particle energy gap, which is determined by $2\min(\Delta, \phi)$, is plotted in the black dashed line.
In proximity of a phase transition, the in-gap mode approaches the $2\Delta/2\phi$ Higgs mode of the vanishing order.\label{fig:collective}}
\end{figure}

The temperature dependence of the mode's frequency is non-trivial -- initially it grows, but then reaches a maximum and goes down with the decrease of either $\phi$ or $\Delta$.
Depending on the shape of the coexistence region, this mode either survives all the way down to $T=0$ or it vanishes at the second transition to a single-order-parameter phase.
This behavior can be seen in Fig.~\ref{fig:collective}.
Note that near the phase transitions, this mode approaches the $2\phi/2\Delta$ amplitude mode of the order that vanishes at that temperature, which is the reason for the softening of the mode in the vicinity of these points.

At the onset of the coexistent phase, the other ($2\phi$) mode is pushed to higher energies, enters the quasiparticle continuum, and quickly becomes overdamped. Thus, it is outside the region of validity of our method of finding $\omega$, and so we do not track it.

\subsection{Damping from antinodal quasiparticles}
\label{sec:antinode}
\begin{figure}
\centering
\includegraphics[width=\linewidth]{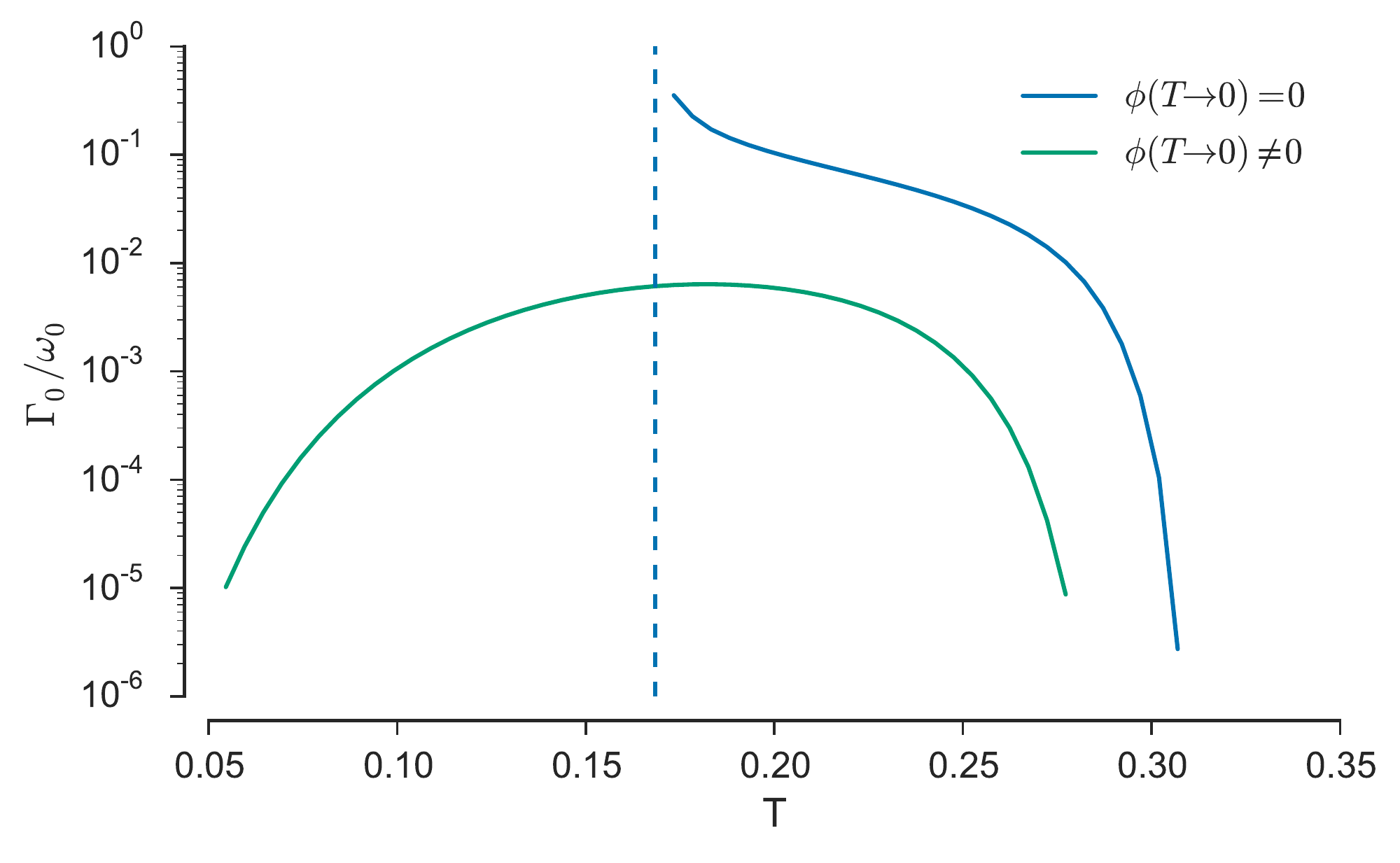}
\caption{(Color online) Damping rate $\Gamma_0$ of the in-gap collective mode in the long wavelength limit for two different values of $V$ corresponding to the two different trajectories depicted in Fig.~\ref{fig:phase}.
Damping is an order of magnitude smaller in the case where $\phi(T \to 0) \neq 0$, and is exponentially suppressed at low temperature due a lack of thermally excited quasiparticles.
In both cases, the decay rate is strongly suppressed in the vicinity of the superconducting $T_c$.
The dashed vertical line indicates the boundary between the coexistent and pure superconductivity phases for the case of the blue curve (c.f. the upper plot of Fig.~\ref{fig:collective}).}
\label{fig:damping}
\end{figure}

\begin{figure}
\centering
    \subfloat[\label{fig:bands_bdw}]{\includegraphics[width=0.9\linewidth]{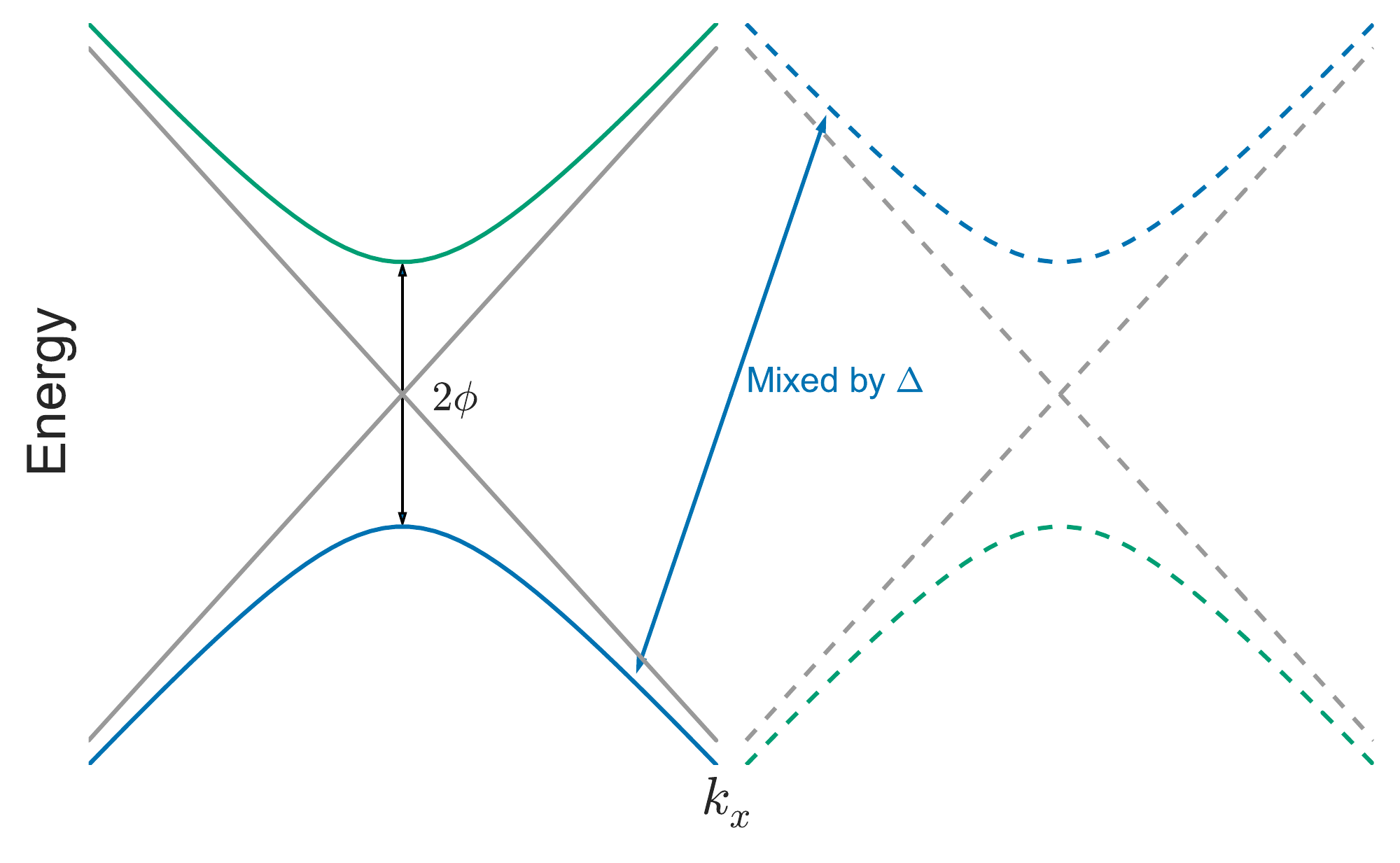}}\\
    \subfloat[\label{fig:bands_sc}]{\includegraphics[width=0.9\linewidth]{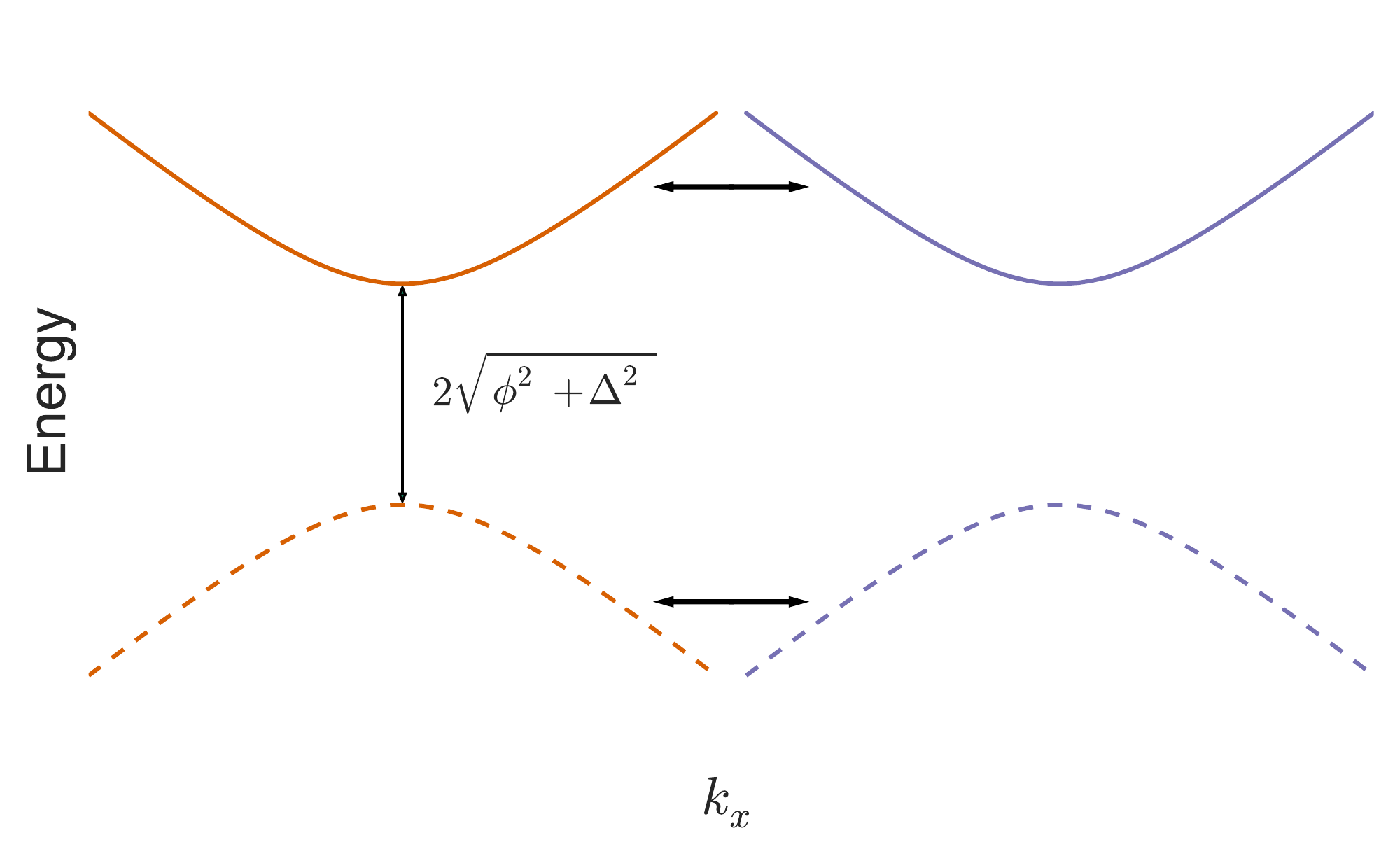}}
\caption{(Color online) Slices of the quasiparticle band structure for $k_y = 0$. Dashed lines indicate the particle hole conjugate of the band of the same color.
(a) The normal state (gray) and BDW state (green/blue) dispersions.
Only transitions between the two solid/dashed bands contribute to $\text{Im}Q^R_{\phi\phi}$.
The onset of superconductivity hybridizes the green/blue bands with their particle-hole conjugates.
(b) Bogoliubov band dispersions in the coexistent state.
Transitions between all bands may contribute to $\text{Im}Q^R_{\phi\phi}$ leading to damping of the Higgs modes (even for those with mass less than $2\Delta$.).
Processes indicated by horizontal arrows are of particular importance as they occupy a finite phase space for arbitrary frequencies within the gap.
\label{fig:bands}}
\end{figure}

As explained in Sec.~\ref{sec:energygap}, the damping rate $\Gamma_0$, can be obtained by expanding the eigenvalues of Eq.~\ref{eq:eigval} about the real part of the zero momentum dispersion $\omega_0$.
The temperature dependence of this damping rate is shown in Fig.~\ref{fig:damping}.
Although the in-gap mode stays below the $(2 \Delta, 2 \phi)$ threshold, its frequency has a finite damping rate, which, furthermore, initially {\it increases} as temperature goes down.
This unusual behavior of the damping arises from the BDW bubble $Q^R_{\phi\phi}$; when just charge order is present, the only scattering which could lead to damping requires at least energy $2\phi$ (as can be seen in Fig.~\ref{fig:bands_bdw}).
All other types of scattering have zero matrix element, and thus there is no damping at $q=0$ for $\omega_0 < 2\phi$.
However, as soon as $\Delta$ becomes non-zero the bands are reconstructed due to hybridization of the BDW bands with their corresponding hole bands, and simultaneously scattering matrix elements between all bands become non-zero, allowing transitions between any two bands to contribute (c.f. Fig.~\ref{fig:bands_sc}).
As a result, there now exist transitions for arbitrarily small frequency (between the two particle/hole bands), giving rise to the damping of collective modes within the gap.

The specific temperature dependence of the damping results from a combination of two effects.
Because we are considering energies $\omega_0 < 2\min(\phi, \Delta)$, we see that transitions from a particle to a hole band (or vice versa) cannot contribute as they will always have energy equal or greater than $ 2\Delta$. Thus, damping must be solely due to scattering between the hole or particle bands.
As $\phi$ decreases, the two particle (and correspondingly the two hole) bands become more similar (in the limit $\phi\to 0$ they are degenerate), increasing the phase space for low energy transitions and therefore leading to greater damping of the BDW amplitude mode.
This in turn leads to an increased damping of the mixed mode, which is visible in Fig.~\ref{fig:damping}.
However, in opposition to this effect, as $\phi \to 0$, the matrix element for scattering between these bands will begin to vanish, as it is proportional to $\phi$.
At some point this second effect will overcome the increase due to the larger phase space, leading to a disappearance of the damping as we approach the critical point at which the charge order disappears.

The competition between scattering elements and band structure generically leads to a non-monotonic temperature dependence of the damping, which in turn means that there exists a region of maximal damping away from which the decay rate remains weak (within the gap).
In the case with $\phi(T\to0) \neq 0$, the BDW order remains sufficiently large that the system never approaches this region of larger damping and thus the decay rate is noticeably smaller than for $\phi(T\to 0) = 0$.
In all cases where the mixed phase exists down to $T=0$, this damping term will be exponentially suppressed at low temperatures as there are no thermally excited quasiparticles available to scatter.

\section{Damping from nodal quasiparticles}
\label{secGL}
The hot-spot model we have used so far is only defined in the antinodal regions, and thus completely ignores the gapless degrees of freedom existing close to the nodes. These can have a particularly strong effect on the damping of the collective modes by providing a low-energy decay channel. However, the contribution of these quasiparticles is different for the different orders. 
We expect the charge order to couple only weakly to the nodal quasiparticles, due to the mismatch between its wavevector $(Q, Q)$ and the wavevector separating the nodes\cite{Vojta} [note that the same argument applies to BDW with $(Q, 0)$ or $(0, Q)$ wavevector]. There is no such restriction for the superconductivity, however, and its amplitude fluctuations are unavoidably damped by the nodal excitations.
To include these effects and to study their consequences for the collective modes, we supplement the  calculation from the previous section with a phenomenological time-dependent Ginzburg-Landau theory. In addition to the more familiar quadratic and quartic in the order parameters terms, this theory contains also first and second derivatives in (real) time. The time-dependent Ginsburg-Landau equations can be written in the following form:\footnote{In general, there is no simple time-dependent extension of Ginzburg-Landau theory, precisely due to the presence of damping, which introduces non-analytic terms (see, for example, I. J. R. Aitchison, G. Metikas, and D. J. Lee,
Phys. Rev. B 62, 6638 (2000), and references therein). We circumvent this difficulty by considering only the $q=0$ limit.}
\begin{equation}
\begin{gathered}
-\frac{\partial^2 \Delta}{\partial t^2}- \gamma_{\Delta} \frac{\partial \Delta}{\partial t}=\frac{\partial \mathcal{F}_{GL}}{\partial \Delta^*},\\
-\frac{\partial^2 \phi}{\partial t^2}- \gamma_{\phi} \frac{\partial \phi}{\partial t}=\frac{\partial \mathcal{F}_{GL}}{\partial \phi^*},
\end{gathered}
\label{TDGL}
\end{equation}
where 
the Ginsburg-Landau action is given by:
\begin{multline}
 \mathcal{F}_{GL}= \alpha_{\phi}|\phi|^2 + \alpha_{\Delta}|\Delta|^2 + \beta_{\phi}|\phi|^4 + \beta_{\Delta}|\Delta|^4  + u|\phi|^2 |\Delta|^2. 
 \nonumber
\end{multline}
The quadratic coefficients $\alpha$ have the usual linear-in-temperature dependence, whereas $\beta$ and $u$ (which parametrises the competition between the two orders) are temperature-independent.\footnote{
We can also add coupling to the lattice degrees of freedom, by including bi-linear terms like $g_{\text{ep}} \phi b$ and $g_{\Delta } \Delta b$, where $b$ is a phonon mode, and $g_{\text{ep}}$ and $g_{\Delta } $ are coupling constants.~\cite{Schaefer2014}
However, the effects of these couplings appear modest (see the appendix), so we will not include them.} Note that expressions for the coefficients in $\mathcal{F}_{GL}$ can be straightforwardly derived from the microscopic theory presented in the previous section\cite{Abrahams} (spatial derivative terms are not included since we are considering only uniform states). 
Although the Ginzburg-Landau theory is strictly applicable only close to the critical region, it can be used beyond its region of validity as an effective model for the collective modes of the system.~\cite{PekkerVarma} For this reason we keep the second-order time derivative terms, which are usually omitted close to the critical temperature.~\cite{larkin2005theory}

The coefficients $\gamma_{\phi}$ and $\gamma_{\Delta}$ are responsible for the damping of the collective modes. It is important to note that despite the symmetric way these terms enter Eq. \ref{TDGL}, they encode very different physics. 
The $\gamma_{\phi}$ term is native to the hot-spot regions. At low energies it is proportional to $\Delta$, since it is only allowed by the band reconstruction (see the discussion in the previous section), whereas above $2\min(\Delta, \phi)$ we can treat it as a constant, originating from the coupling of the fluctuations to the high-energy quasiparticle continuum. In contrast, the main contribution to the $\gamma_{\Delta}$ term originates from the nodal regions (and thus is completely absent in the hot-spot-only approach of the previous section).  Close to $T_c$ we can obtain its temperature dependence from the following qualitative considerations. This term is proportional to the number of available states at the oscillation frequency, given by $\sim \rho (\omega) \tanh{(\omega/4 T)}$.~\cite{Sharapov} Linearizing the density of states close to the nodes $\rho(\omega) \sim \omega$, and approximating the frequency as $\omega \approx 2 \Delta$ we finally get for the damping terms of the slow mode
\begin{equation}
\gamma_{\Delta} \approx \gamma^0_{\Delta} \Delta^2 = \gamma^0_{\Delta} (T_c - T), \ \gamma_{\phi} \approx \gamma^0_{\phi} \Delta = \gamma^0_{\phi} \sqrt{T_c - T}\nonumber
\end{equation}
(we have expanded in powers of $\Delta$). Note that we have thus determined the temperature dependence of $\gamma_{\phi}$ and $\gamma_{\Delta}$, but their relative strength at some fixed temperature depends on the parameters of the microscopic models (like $V$), which cannot be estimated within our phenomenological theory. However, given the general temperature dependence of  $\gamma_{\phi}$ and $\gamma_{ \Delta}$, we expect the antinodal particles to dominate damping sufficiently close to $T_c$ ($\Delta$ vs. $\Delta^2$), whereas at low temperatures the nodal excitations take over -- $\gamma_{\Delta}$ stays finite for $T \rightarrow 0$, while $\gamma_{\phi}$ goes to zero exponentially.

\begin{figure}[h]
\centering
\includegraphics[width=0.95\linewidth]{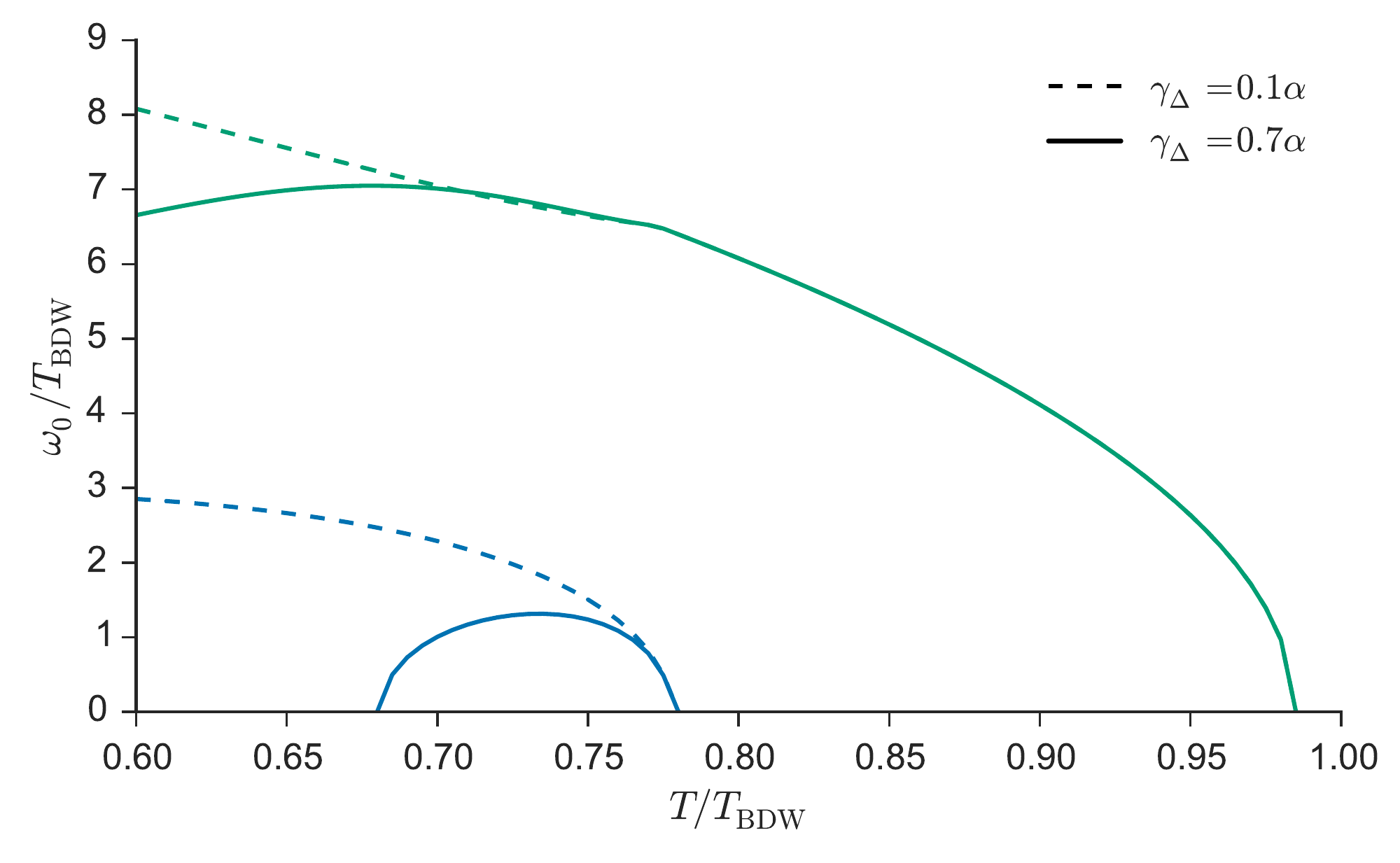}\\
\includegraphics[width=0.95\linewidth]{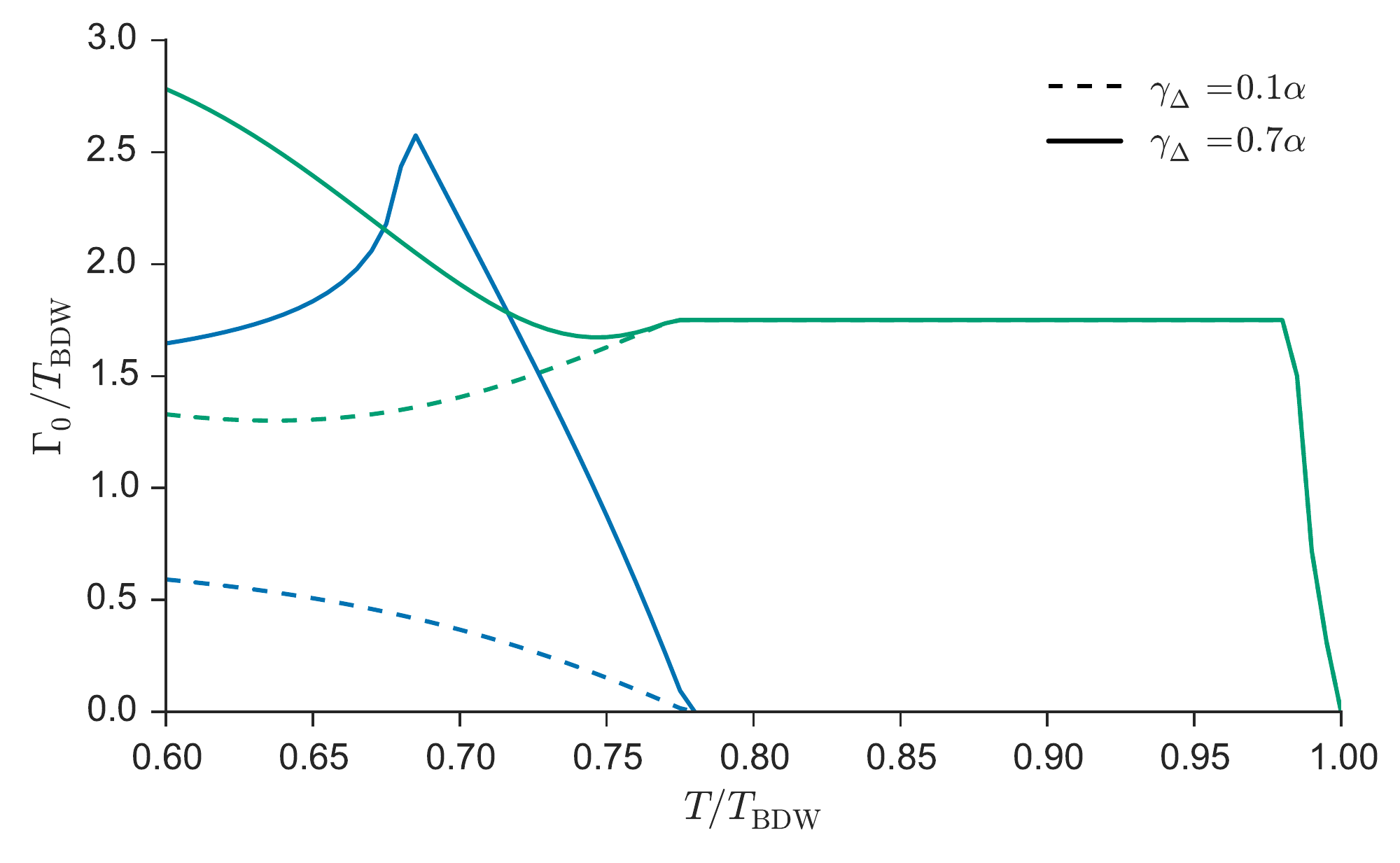}
\caption{(Color online) The evolution of $\omega_0$ (top panel) and $\Gamma_0$ (bottom panel) of the two Higgs modes with temperature.  For each mode the cases of weak and strong damping from the nodal regions are shown [$\gamma_{\Delta} = 0.1 \alpha$ (dashed line) and $\gamma_{\Delta} = 0.7 \alpha$ (solid line), respectively].
$\gamma_{\phi}$, the damping from antinodal region, is the same on both plots.
\label{fig:glplot}}
\end{figure}

To obtain the frequencies and damping of the mixed modes we expand $\phi(t)$ and $\Delta(t)$ around the mean field values of the order parameters $\phi_0$ and $\Delta_0$: $\phi(t) = \phi_0 + \delta \phi(t)$ and $\Delta(t) = \Delta_0 + \delta \Delta(t)$. Assuming that $\delta \phi(t)$ and $\delta \Delta(t)$ are relatively small we can simplify  Eq.~\ref{TDGL} by keeping only the terms linear in $\delta \phi$ and $\delta \Delta$. Since we are interested in the collective modes we write their time dependence as $ e^{- i \omega t}$.  Inserting this ansatz in the linearized equations, we can exclude $\delta \phi$ and $\delta \Delta$ altogether, and finally arrive at the following equation for $\omega$:
\begin{multline}
 \omega^2 + i \gamma_{\phi} \omega + 2 (\alpha_{\phi} + u \Delta_0^2) - \frac{(2 u \phi_0 \Delta_0)^2}{2(\alpha_{\Delta} + u \phi_0^2) + i  \gamma_{\Delta} \omega  +  \omega^2}=0.\\
 \label{GL_freq}
\end{multline}
We solve it numerically (with $\phi_0(T)$ and $\Delta_0(T)$ determined by the time-independent mean-field equations), and obtain both complex and purely imaginary solutions for $\omega$. The former solutions are oscillatory (with Re$[\omega]$ giving the frequency of the uniform oscillations around the mean field values), while the latter represent exponential decay.
We show the real and the imaginary parts of the two $\omega$ solutions  as a function of temperature in Fig. \ref{fig:glplot}. There we plot $\omega_0$ and $\Gamma_0$ for two different strengths of $\gamma^0_{\Delta}$, as a comparison between small and large contribution from the nodal quasiparticles, respectively. For small $\gamma^0_{\Delta}$
we can see that both the real and imaginary parts of the frequency of the hybridized modes show behavior similar to that obtained in the previous section. However, when we increase $\gamma^0_{\Delta}$ we see not only enhancement of the damping of both modes, but also decrease of their real frequencies (the top panel of Fig. \ref{fig:glplot}). Although the effect is more dramatic for the in-gap mode, which now exists only in a narrow region below $T_c$, it is significant for the fast one as well. This is a consequence of one important feature of Eq. \ref{GL_freq} -- the coupling of the two channels mixes their real and imaginary parts. Thus, increase of the damping leads to the gradual suppression of the real part of both mixed modes. Note also that the disappearance of $\omega_0$ of the in-gap mode corresponds to a peak in its $\Gamma_0$. 

\section{Discussion and Conclusion}

Note that our calculation is to some extent complementary to those in Refs. \onlinecite{Hung2014, Moor2014}. These works studied the dynamics of the system after an external perturbation, and were done in the time domain, thus allowing direct comparison with the experimental data. The temperature dependence of the frequencies  extracted in Ref. \onlinecite{Hung2014} appears consistent
with our calculation, as it shows a low-frequency mode appearing below the superconducting transition.

The experiments have not observed a soft mode close to either charge or superconducting transition temperatures. Instead, the frequency of the identified amplitude mode stays almost constant, with only a small decrease in frequency at the superconducting $T_c$ observed in Ref. \onlinecite{Hinton2013}, and no clear change seen in Ref. \onlinecite{Torchinsky2013}. This appears consistent with the behavior of the high-energy mode in the case of strong damping from the nodal regions (see section \ref{secGL}). This damping can effectively ``squeeze" the low-frequency mode inside a very narrow region close to $T_c$ (where it would be difficult to observe), and could also lead to the decrease of the frequency of the fast mode, observed in Ref. \onlinecite{Hinton2013} (note that a different phenomenological explanation for this decrease, based on time-independent Ginzburg-Landau theory, was given in Ref. \onlinecite{Hinton2013}). In contrast, the absence of softening close to $T_{BDW}$ appears incompatible with our calculation, and requires alternative explanations (such as optical phonons).\cite{Hinton2013} 

In conclusion, we have studied the collective modes for the bond density wave and superconducting order parameters expected to exist in the pseudogap state of cuprates. In the pure BDW phase we observed the conventional amplitude mode with frequency starting at $2 \phi$. In the coexistent phase two collective modes representing the coupled oscillations of the amplitudes of the order parameters are present. One of them is soft at the superconducting critical temperature, and despite having frequency $\omega_0 < 2\min(\phi, \Delta)$ is (weakly) damped, due to band-structure reconstruction caused by superconductivity. The other mixed mode is continuously connected to the pure BDW mode, with frequency pushed up in the coexisting regime.  To study the effects of damping originating from the nodal regions, we developed a phenomenological time-dependent Ginzburg-Landau theory. We demonstrated that strong damping can have significant effect on the real frequency of the modes.

\begin{acknowledgements}
We are grateful to D.H. Torchinsky for enlightening discussion.
This work was supported by U.S. Department of Energy {BES-DESC0001911} and Simons Foundation.
\end{acknowledgements}

\appendix*
\section{Effect of phonons}

Beyond just the non-retarded interaction considered above, one can also consider the effect of phonons on the collective modes.
Here we will take this into account by considering the contribution of the frequency dependent phonon-mediated interaction between electrons to the collective mode propagators.
In particular, we will project this interaction onto a hot spot model by taking the phonon momentum to be the fixed wavevector $\vec Q$ separating the hot spots which are being paired -- this is the same approximation that one uses on the non-retarded interaction in deriving the hot spot model.

A simple $A_{1g}$ symmetry phonon has no effect on the collective modes due to the pure $d$-wave symmetry of the order parameters.
However, in reality we expect some direct order parameter-phonon coupling, either because there is a phonon mode with the correct symmetry ($B_{1g}$), or because in real systems the order parameter would not necessarily have a pure $d$-wave symmetry, but could have an $s$-wave component admixed.
Regardless of the exact nature of the coupling, it gives rise to a term in the mean field theory which includes the phonon-mediated interaction as
\begin{multline}
H_{ph} = f\sum_{k, \epsilon_n, \omega_m} U(\omega_m)\\
\times
    \left(\phi(\omega_m) c^\dagger_{1\sigma}(k, \epsilon_n - \omega_m)c_{2\sigma}(k, \epsilon_n) + h.c.\right)
\end{multline}
where
\[U(\omega_m) = \frac{g^2_\text{ep}}{2} \frac{\Omega_Q}{\omega_m^2 + \Omega_Q^2}\]
is an Einstein phonon type propagator and $f$ is a constant of order one arising from the form factor of the electron-phonon vertex.

\begin{figure}
\centering
\vspace{10pt}
\includegraphics[width=\linewidth]{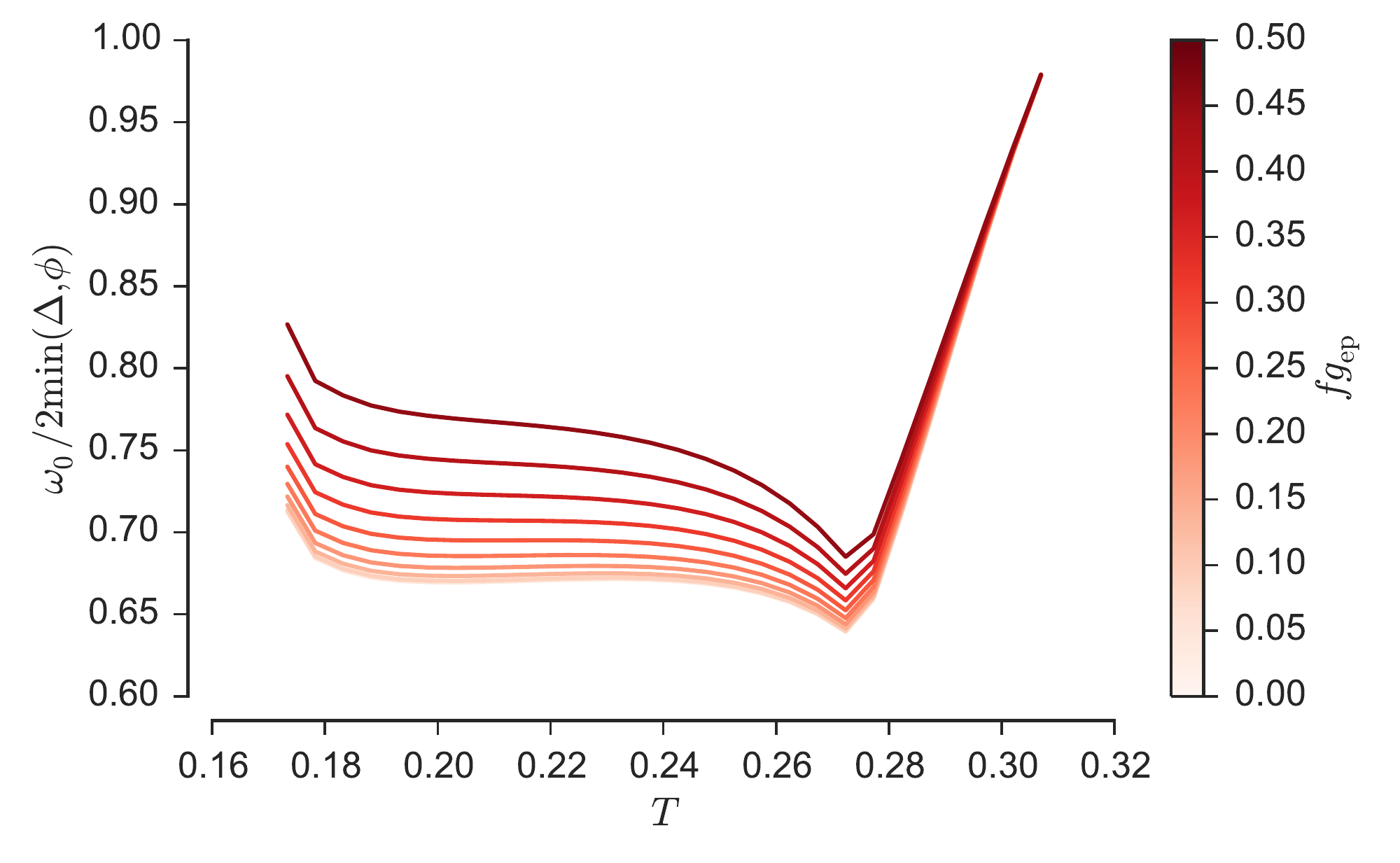}
\caption{(Color online) The in-gap collective mode mass $\omega_0$ (relative to the minimum pair creation energy) as a function of the electron phonon coupling $g$, for $\Omega_{Q} = 1.$.
The $g=0$ line reproduces the behavior shown in Fig.~\ref{fig:collective}.}
\label{fig:phonon_coupling}
\end{figure}

If we consider the effect of this term on the charge collective mode, we find that it can be captured by the replacement $g_c \to \tilde g_c(i \omega_m)$.
We absorb the $\omega=0$ component into the definition of $g_c$ (as that is what determines the static mean-field solution) and include the remaining frequency dependent part in our calculation of the collective modes.
Upon analytic continuation to real frequency, this amounts to the substitution
\begin{equation}
g_c \to \tilde g_c(\omega) = g_c - \frac{f g^2_\text{ep}}{\Omega} - f g_\text{ep}^2 \frac{\Omega}{\omega^2 - \Omega^2}
\end{equation}
in the collective mode equations (the additive constant is chosen so that we recover $\tilde g_c(\omega=0) = g_c$).
The previous analysis can now be repeated for a range of electron phonon couplings.
As can be seen in Fig.~\ref{fig:phonon_coupling}, the coupling to phonons tends to push the collective mode gap downward, while leaving the softening at the phase transitions unmodified.
Overall, the qualitative behavior of the mode is not markedly different.

\bibliography{references}

\begin{thebibliography}{44}%
\makeatletter
\providecommand \@ifxundefined [1]{%
 \@ifx{#1\undefined}
}%
\providecommand \@ifnum [1]{%
 \ifnum #1\expandafter \@firstoftwo
 \else \expandafter \@secondoftwo
 \fi
}%
\providecommand \@ifx [1]{%
 \ifx #1\expandafter \@firstoftwo
 \else \expandafter \@secondoftwo
 \fi
}%
\providecommand \natexlab [1]{#1}%
\providecommand \enquote  [1]{``#1''}%
\providecommand \bibnamefont  [1]{#1}%
\providecommand \bibfnamefont [1]{#1}%
\providecommand \citenamefont [1]{#1}%
\providecommand \href@noop [0]{\@secondoftwo}%
\providecommand \href [0]{\begingroup \@sanitize@url \@href}%
\providecommand \@href[1]{\@@startlink{#1}\@@href}%
\providecommand \@@href[1]{\endgroup#1\@@endlink}%
\providecommand \@sanitize@url [0]{\catcode `\\12\catcode `\$12\catcode
  `\&12\catcode `\#12\catcode `\^12\catcode `\_12\catcode `\%12\relax}%
\providecommand \@@startlink[1]{}%
\providecommand \@@endlink[0]{}%
\providecommand \url  [0]{\begingroup\@sanitize@url \@url }%
\providecommand \@url [1]{\endgroup\@href {#1}{\urlprefix }}%
\providecommand \urlprefix  [0]{URL }%
\providecommand \Eprint [0]{\href }%
\providecommand \doibase [0]{http://dx.doi.org/}%
\providecommand \selectlanguage [0]{\@gobble}%
\providecommand \bibinfo  [0]{\@secondoftwo}%
\providecommand \bibfield  [0]{\@secondoftwo}%
\providecommand \translation [1]{[#1]}%
\providecommand \BibitemOpen [0]{}%
\providecommand \bibitemStop [0]{}%
\providecommand \bibitemNoStop [0]{.\EOS\space}%
\providecommand \EOS [0]{\spacefactor3000\relax}%
\providecommand \BibitemShut  [1]{\csname bibitem#1\endcsname}%
\let\auto@bib@innerbib\@empty
\bibitem [{\citenamefont {Taillefer}(2010)}]{Taillefer2010a}%
  \BibitemOpen
  \bibfield  {author} {\bibinfo {author} {\bibfnamefont {L.}~\bibnamefont
  {Taillefer}},\ }\href {\doibase 10.1146/annurev-conmatphys-070909-104117}
  {\bibfield  {journal} {\bibinfo  {journal} {Annu. Rev. Condens. Matter
  Phys.}\ }\textbf {\bibinfo {volume} {1}},\ \bibinfo {pages} {51} (\bibinfo
  {year} {2010})}\BibitemShut {NoStop}%
\bibitem [{\citenamefont {Keimer}\ \emph {et~al.}(2015)\citenamefont {Keimer},
  \citenamefont {Kivelson}, \citenamefont {Norman}, \citenamefont {Uchida},\
  and\ \citenamefont {Zaanen}}]{Keimer2014}%
  \BibitemOpen
  \bibfield  {author} {\bibinfo {author} {\bibfnamefont {B.}~\bibnamefont
  {Keimer}}, \bibinfo {author} {\bibfnamefont {S.~a.}\ \bibnamefont
  {Kivelson}}, \bibinfo {author} {\bibfnamefont {M.~R.}\ \bibnamefont
  {Norman}}, \bibinfo {author} {\bibfnamefont {S.}~\bibnamefont {Uchida}}, \
  and\ \bibinfo {author} {\bibfnamefont {J.}~\bibnamefont {Zaanen}},\ }\href
  {\doibase 10.1038/nature14165} {\bibfield  {journal} {\bibinfo  {journal}
  {Nature}\ }\textbf {\bibinfo {volume} {518}},\ \bibinfo {pages} {179}
  (\bibinfo {year} {2015})}\BibitemShut {NoStop}%
\bibitem [{\citenamefont {Ghiringhelli}\ \emph {et~al.}(2012)\citenamefont
  {Ghiringhelli}, \citenamefont {Le~Tacon}, \citenamefont {Minola},
  \citenamefont {Blanco-Canosa}, \citenamefont {Mazzoli}, \citenamefont
  {Brookes}, \citenamefont {De~Luca}, \citenamefont {Frano}, \citenamefont
  {Hawthorn}, \citenamefont {He}, \citenamefont {Loew}, \citenamefont {Sala},
  \citenamefont {Peets}, \citenamefont {Salluzzo}, \citenamefont {Schierle},
  \citenamefont {Sutarto}, \citenamefont {Sawatzky}, \citenamefont {Weschke},
  \citenamefont {Keimer},\ and\ \citenamefont {Braicovich}}]{Ghiringhelli2012}%
  \BibitemOpen
  \bibfield  {author} {\bibinfo {author} {\bibfnamefont {G.}~\bibnamefont
  {Ghiringhelli}}, \bibinfo {author} {\bibfnamefont {M.}~\bibnamefont
  {Le~Tacon}}, \bibinfo {author} {\bibfnamefont {M.}~\bibnamefont {Minola}},
  \bibinfo {author} {\bibfnamefont {S.}~\bibnamefont {Blanco-Canosa}}, \bibinfo
  {author} {\bibfnamefont {C.}~\bibnamefont {Mazzoli}}, \bibinfo {author}
  {\bibfnamefont {N.~B.}\ \bibnamefont {Brookes}}, \bibinfo {author}
  {\bibfnamefont {G.~M.}\ \bibnamefont {De~Luca}}, \bibinfo {author}
  {\bibfnamefont {A.}~\bibnamefont {Frano}}, \bibinfo {author} {\bibfnamefont
  {D.~G.}\ \bibnamefont {Hawthorn}}, \bibinfo {author} {\bibfnamefont
  {F.}~\bibnamefont {He}}, \bibinfo {author} {\bibfnamefont {T.}~\bibnamefont
  {Loew}}, \bibinfo {author} {\bibfnamefont {M.~M.}\ \bibnamefont {Sala}},
  \bibinfo {author} {\bibfnamefont {D.~C.}\ \bibnamefont {Peets}}, \bibinfo
  {author} {\bibfnamefont {M.}~\bibnamefont {Salluzzo}}, \bibinfo {author}
  {\bibfnamefont {E.}~\bibnamefont {Schierle}}, \bibinfo {author}
  {\bibfnamefont {R.}~\bibnamefont {Sutarto}}, \bibinfo {author} {\bibfnamefont
  {G.~A.}\ \bibnamefont {Sawatzky}}, \bibinfo {author} {\bibfnamefont
  {E.}~\bibnamefont {Weschke}}, \bibinfo {author} {\bibfnamefont
  {B.}~\bibnamefont {Keimer}}, \ and\ \bibinfo {author} {\bibfnamefont
  {L.}~\bibnamefont {Braicovich}},\ }\href {\doibase 10.1126/science.1223532}
  {\bibfield  {journal} {\bibinfo  {journal} {Science}\ }\textbf {\bibinfo
  {volume} {337}},\ \bibinfo {pages} {821} (\bibinfo {year}
  {2012})}\BibitemShut {NoStop}%
\bibitem [{\citenamefont {Chang}\ \emph {et~al.}(2012)\citenamefont {Chang},
  \citenamefont {Blackburn}, \citenamefont {Holmes}, \citenamefont
  {Christensen}, \citenamefont {Larsen}, \citenamefont {Mesot}, \citenamefont
  {Liang}, \citenamefont {Bonn}, \citenamefont {Hardy}, \citenamefont
  {Watenphul}, \citenamefont {Zimmermann}, \citenamefont {Forgan},\ and\
  \citenamefont {Hayden}}]{Chang2012}%
  \BibitemOpen
  \bibfield  {author} {\bibinfo {author} {\bibfnamefont {J.}~\bibnamefont
  {Chang}}, \bibinfo {author} {\bibfnamefont {E.}~\bibnamefont {Blackburn}},
  \bibinfo {author} {\bibfnamefont {A.~T.}\ \bibnamefont {Holmes}}, \bibinfo
  {author} {\bibfnamefont {N.~B.}\ \bibnamefont {Christensen}}, \bibinfo
  {author} {\bibfnamefont {J.}~\bibnamefont {Larsen}}, \bibinfo {author}
  {\bibfnamefont {J.}~\bibnamefont {Mesot}}, \bibinfo {author} {\bibfnamefont
  {R.}~\bibnamefont {Liang}}, \bibinfo {author} {\bibfnamefont {D.~A.}\
  \bibnamefont {Bonn}}, \bibinfo {author} {\bibfnamefont {W.~N.}\ \bibnamefont
  {Hardy}}, \bibinfo {author} {\bibfnamefont {A.}~\bibnamefont {Watenphul}},
  \bibinfo {author} {\bibfnamefont {M.~V.}\ \bibnamefont {Zimmermann}},
  \bibinfo {author} {\bibfnamefont {E.~M.}\ \bibnamefont {Forgan}}, \ and\
  \bibinfo {author} {\bibfnamefont {S.~M.}\ \bibnamefont {Hayden}},\ }\href
  {\doibase 10.1038/nphys2456} {\bibfield  {journal} {\bibinfo  {journal} {Nat.
  Phys.}\ }\textbf {\bibinfo {volume} {8}},\ \bibinfo {pages} {871} (\bibinfo
  {year} {2012})}\BibitemShut {NoStop}%
\bibitem [{\citenamefont {Achkar}\ \emph {et~al.}(2012)\citenamefont {Achkar},
  \citenamefont {Sutarto}, \citenamefont {Mao}, \citenamefont {He},
  \citenamefont {Frano}, \citenamefont {Blanco-Canosa}, \citenamefont
  {Le~Tacon}, \citenamefont {Ghiringhelli}, \citenamefont {Braicovich},
  \citenamefont {Minola}, \citenamefont {Moretti~Sala}, \citenamefont
  {Mazzoli}, \citenamefont {Liang}, \citenamefont {Bonn}, \citenamefont
  {Hardy}, \citenamefont {Keimer}, \citenamefont {Sawatzky},\ and\
  \citenamefont {Hawthorn}}]{Achkar2012}%
  \BibitemOpen
  \bibfield  {author} {\bibinfo {author} {\bibfnamefont {A.~J.}\ \bibnamefont
  {Achkar}}, \bibinfo {author} {\bibfnamefont {R.}~\bibnamefont {Sutarto}},
  \bibinfo {author} {\bibfnamefont {X.}~\bibnamefont {Mao}}, \bibinfo {author}
  {\bibfnamefont {F.}~\bibnamefont {He}}, \bibinfo {author} {\bibfnamefont
  {A.}~\bibnamefont {Frano}}, \bibinfo {author} {\bibfnamefont
  {S.}~\bibnamefont {Blanco-Canosa}}, \bibinfo {author} {\bibfnamefont
  {M.}~\bibnamefont {Le~Tacon}}, \bibinfo {author} {\bibfnamefont
  {G.}~\bibnamefont {Ghiringhelli}}, \bibinfo {author} {\bibfnamefont
  {L.}~\bibnamefont {Braicovich}}, \bibinfo {author} {\bibfnamefont
  {M.}~\bibnamefont {Minola}}, \bibinfo {author} {\bibfnamefont
  {M.}~\bibnamefont {Moretti~Sala}}, \bibinfo {author} {\bibfnamefont
  {C.}~\bibnamefont {Mazzoli}}, \bibinfo {author} {\bibfnamefont
  {R.}~\bibnamefont {Liang}}, \bibinfo {author} {\bibfnamefont {D.~A.}\
  \bibnamefont {Bonn}}, \bibinfo {author} {\bibfnamefont {W.~N.}\ \bibnamefont
  {Hardy}}, \bibinfo {author} {\bibfnamefont {B.}~\bibnamefont {Keimer}},
  \bibinfo {author} {\bibfnamefont {G.~A.}\ \bibnamefont {Sawatzky}}, \ and\
  \bibinfo {author} {\bibfnamefont {D.~G.}\ \bibnamefont {Hawthorn}},\ }\href
  {\doibase 10.1103/PhysRevLett.109.167001} {\bibfield  {journal} {\bibinfo
  {journal} {Phys. Rev. Lett.}\ }\textbf {\bibinfo {volume} {109}},\ \bibinfo
  {pages} {167001} (\bibinfo {year} {2012})}\BibitemShut {NoStop}%
\bibitem [{\citenamefont {Coslovich}\ \emph {et~al.}(2013)\citenamefont
  {Coslovich}, \citenamefont {Giannetti}, \citenamefont {Cilento},
  \citenamefont {Dal~Conte}, \citenamefont {Abebaw}, \citenamefont {Bossini},
  \citenamefont {Ferrini}, \citenamefont {Eisaki}, \citenamefont {Greven},
  \citenamefont {Damascelli},\ and\ \citenamefont
  {Parmigiani}}]{Coslovich2013}%
  \BibitemOpen
  \bibfield  {author} {\bibinfo {author} {\bibfnamefont {G.}~\bibnamefont
  {Coslovich}}, \bibinfo {author} {\bibfnamefont {C.}~\bibnamefont
  {Giannetti}}, \bibinfo {author} {\bibfnamefont {F.}~\bibnamefont {Cilento}},
  \bibinfo {author} {\bibfnamefont {S.}~\bibnamefont {Dal~Conte}}, \bibinfo
  {author} {\bibfnamefont {T.}~\bibnamefont {Abebaw}}, \bibinfo {author}
  {\bibfnamefont {D.}~\bibnamefont {Bossini}}, \bibinfo {author} {\bibfnamefont
  {G.}~\bibnamefont {Ferrini}}, \bibinfo {author} {\bibfnamefont
  {H.}~\bibnamefont {Eisaki}}, \bibinfo {author} {\bibfnamefont
  {M.}~\bibnamefont {Greven}}, \bibinfo {author} {\bibfnamefont
  {A.}~\bibnamefont {Damascelli}}, \ and\ \bibinfo {author} {\bibfnamefont
  {F.}~\bibnamefont {Parmigiani}},\ }\href {\doibase
  10.1103/PhysRevLett.110.107003} {\bibfield  {journal} {\bibinfo  {journal}
  {Phys. Rev. Lett.}\ }\textbf {\bibinfo {volume} {110}},\ \bibinfo {pages}
  {107003} (\bibinfo {year} {2013})}\BibitemShut {NoStop}%
\bibitem [{\citenamefont {Fujita}\ \emph {et~al.}(2014)\citenamefont {Fujita},
  \citenamefont {Hamidian}, \citenamefont {Edkins}, \citenamefont {Kim},
  \citenamefont {Kohsaka}, \citenamefont {Azuma}, \citenamefont {Takano},
  \citenamefont {Takagi}, \citenamefont {Eisaki}, \citenamefont {Uchida},
  \citenamefont {Allais}, \citenamefont {Lawler}, \citenamefont {Kim},
  \citenamefont {Sachdev},\ and\ \citenamefont {Davis}}]{Fujita2014}%
  \BibitemOpen
  \bibfield  {author} {\bibinfo {author} {\bibfnamefont {K.}~\bibnamefont
  {Fujita}}, \bibinfo {author} {\bibfnamefont {M.~H.}\ \bibnamefont
  {Hamidian}}, \bibinfo {author} {\bibfnamefont {S.~D.}\ \bibnamefont
  {Edkins}}, \bibinfo {author} {\bibfnamefont {C.~K.}\ \bibnamefont {Kim}},
  \bibinfo {author} {\bibfnamefont {Y.}~\bibnamefont {Kohsaka}}, \bibinfo
  {author} {\bibfnamefont {M.}~\bibnamefont {Azuma}}, \bibinfo {author}
  {\bibfnamefont {M.}~\bibnamefont {Takano}}, \bibinfo {author} {\bibfnamefont
  {H.}~\bibnamefont {Takagi}}, \bibinfo {author} {\bibfnamefont
  {H.}~\bibnamefont {Eisaki}}, \bibinfo {author} {\bibfnamefont {S.-i.}\
  \bibnamefont {Uchida}}, \bibinfo {author} {\bibfnamefont {A.}~\bibnamefont
  {Allais}}, \bibinfo {author} {\bibfnamefont {M.~J.}\ \bibnamefont {Lawler}},
  \bibinfo {author} {\bibfnamefont {E.-A.}\ \bibnamefont {Kim}}, \bibinfo
  {author} {\bibfnamefont {S.}~\bibnamefont {Sachdev}}, \ and\ \bibinfo
  {author} {\bibfnamefont {J.~C.~S.}\ \bibnamefont {Davis}},\ }\href {\doibase
  10.1073/pnas.1406297111} {\bibfield  {journal} {\bibinfo  {journal} {Proc.
  Natl. Acad. Sci.}\ }\textbf {\bibinfo {volume} {111}},\ \bibinfo {pages}
  {E3026} (\bibinfo {year} {2014})}\BibitemShut {NoStop}%
\bibitem [{\citenamefont {Comin}\ \emph {et~al.}(2014)\citenamefont {Comin},
  \citenamefont {Sutarto}, \citenamefont {He}, \citenamefont {Neto},
  \citenamefont {Chauviere}, \citenamefont {Frano}, \citenamefont {Liang},
  \citenamefont {Hardy}, \citenamefont {Bonn}, \citenamefont {Yoshida},
  \citenamefont {Eisaki}, \citenamefont {Hoffman}, \citenamefont {Keimer},
  \citenamefont {Sawatzky},\ and\ \citenamefont {Damascelli}}]{Comin2014}%
  \BibitemOpen
  \bibfield  {author} {\bibinfo {author} {\bibfnamefont {R.}~\bibnamefont
  {Comin}}, \bibinfo {author} {\bibfnamefont {R.}~\bibnamefont {Sutarto}},
  \bibinfo {author} {\bibfnamefont {F.}~\bibnamefont {He}}, \bibinfo {author}
  {\bibfnamefont {E.~D.~S.}\ \bibnamefont {Neto}}, \bibinfo {author}
  {\bibfnamefont {L.}~\bibnamefont {Chauviere}}, \bibinfo {author}
  {\bibfnamefont {A.}~\bibnamefont {Frano}}, \bibinfo {author} {\bibfnamefont
  {R.}~\bibnamefont {Liang}}, \bibinfo {author} {\bibfnamefont {W.~N.}\
  \bibnamefont {Hardy}}, \bibinfo {author} {\bibfnamefont {D.}~\bibnamefont
  {Bonn}}, \bibinfo {author} {\bibfnamefont {Y.}~\bibnamefont {Yoshida}},
  \bibinfo {author} {\bibfnamefont {H.}~\bibnamefont {Eisaki}}, \bibinfo
  {author} {\bibfnamefont {J.~E.}\ \bibnamefont {Hoffman}}, \bibinfo {author}
  {\bibfnamefont {B.}~\bibnamefont {Keimer}}, \bibinfo {author} {\bibfnamefont
  {G.~a.}\ \bibnamefont {Sawatzky}}, \ and\ \bibinfo {author} {\bibfnamefont
  {A.}~\bibnamefont {Damascelli}},\ }\href {http://arxiv.org/abs/1402.5415} {\
  (\bibinfo {year} {2014})},\ \Eprint {http://arxiv.org/abs/1402.5415}
  {arXiv:1402.5415} \BibitemShut {NoStop}%
\bibitem [{\citenamefont {F\"{o}rst}\ \emph {et~al.}(2014)\citenamefont
  {F\"{o}rst}, \citenamefont {Frano}, \citenamefont {Kaiser}, \citenamefont
  {Mankowsky}, \citenamefont {Hunt}, \citenamefont {Turner}, \citenamefont
  {Dakovski}, \citenamefont {Minitti}, \citenamefont {Robinson}, \citenamefont
  {Loew}, \citenamefont {{Le Tacon}}, \citenamefont {Keimer}, \citenamefont
  {Hill}, \citenamefont {Cavalleri},\ and\ \citenamefont {Dhesi}}]{Forst2014}%
  \BibitemOpen
  \bibfield  {author} {\bibinfo {author} {\bibfnamefont {M.}~\bibnamefont
  {F\"{o}rst}}, \bibinfo {author} {\bibfnamefont {A.}~\bibnamefont {Frano}},
  \bibinfo {author} {\bibfnamefont {S.}~\bibnamefont {Kaiser}}, \bibinfo
  {author} {\bibfnamefont {R.}~\bibnamefont {Mankowsky}}, \bibinfo {author}
  {\bibfnamefont {C.~R.}\ \bibnamefont {Hunt}}, \bibinfo {author}
  {\bibfnamefont {J.~J.}\ \bibnamefont {Turner}}, \bibinfo {author}
  {\bibfnamefont {G.~L.}\ \bibnamefont {Dakovski}}, \bibinfo {author}
  {\bibfnamefont {M.~P.}\ \bibnamefont {Minitti}}, \bibinfo {author}
  {\bibfnamefont {J.}~\bibnamefont {Robinson}}, \bibinfo {author}
  {\bibfnamefont {T.}~\bibnamefont {Loew}}, \bibinfo {author} {\bibfnamefont
  {M.}~\bibnamefont {{Le Tacon}}}, \bibinfo {author} {\bibfnamefont
  {B.}~\bibnamefont {Keimer}}, \bibinfo {author} {\bibfnamefont {J.~P.}\
  \bibnamefont {Hill}}, \bibinfo {author} {\bibfnamefont {A.}~\bibnamefont
  {Cavalleri}}, \ and\ \bibinfo {author} {\bibfnamefont {S.~S.}\ \bibnamefont
  {Dhesi}},\ }\href {\doibase 10.1103/PhysRevB.90.184514} {\bibfield  {journal}
  {\bibinfo  {journal} {Phys. Rev. B}\ }\textbf {\bibinfo {volume} {90}},\
  \bibinfo {pages} {184514} (\bibinfo {year} {2014})}\BibitemShut {NoStop}%
\bibitem [{\citenamefont {Demsar}\ \emph {et~al.}(1999)\citenamefont {Demsar},
  \citenamefont {Biljakovi\ifmmode~\acute{c}\else \'{c}\fi{}},\ and\
  \citenamefont {Mihailovic}}]{Demsar1999}%
  \BibitemOpen
  \bibfield  {author} {\bibinfo {author} {\bibfnamefont {J.}~\bibnamefont
  {Demsar}}, \bibinfo {author} {\bibfnamefont {K.}~\bibnamefont
  {Biljakovi\ifmmode~\acute{c}\else \'{c}\fi{}}}, \ and\ \bibinfo {author}
  {\bibfnamefont {D.}~\bibnamefont {Mihailovic}},\ }\href {\doibase
  10.1103/PhysRevLett.83.800} {\bibfield  {journal} {\bibinfo  {journal} {Phys.
  Rev. Lett.}\ }\textbf {\bibinfo {volume} {83}},\ \bibinfo {pages} {800}
  (\bibinfo {year} {1999})}\BibitemShut {NoStop}%
\bibitem [{\citenamefont {Hinton}\ \emph
  {et~al.}(2013{\natexlab{a}})\citenamefont {Hinton}, \citenamefont {Koralek},
  \citenamefont {Yu}, \citenamefont {Motoyama}, \citenamefont {Lu},
  \citenamefont {Vishwanath}, \citenamefont {Greven},\ and\ \citenamefont
  {Orenstein}}]{HintonPRL}%
  \BibitemOpen
  \bibfield  {author} {\bibinfo {author} {\bibfnamefont {J.~P.}\ \bibnamefont
  {Hinton}}, \bibinfo {author} {\bibfnamefont {J.~D.}\ \bibnamefont {Koralek}},
  \bibinfo {author} {\bibfnamefont {G.}~\bibnamefont {Yu}}, \bibinfo {author}
  {\bibfnamefont {E.~M.}\ \bibnamefont {Motoyama}}, \bibinfo {author}
  {\bibfnamefont {Y.~M.}\ \bibnamefont {Lu}}, \bibinfo {author} {\bibfnamefont
  {A.}~\bibnamefont {Vishwanath}}, \bibinfo {author} {\bibfnamefont
  {M.}~\bibnamefont {Greven}}, \ and\ \bibinfo {author} {\bibfnamefont
  {J.}~\bibnamefont {Orenstein}},\ }\href {\doibase
  10.1103/PhysRevLett.110.217002} {\bibfield  {journal} {\bibinfo  {journal}
  {Phys. Rev. Lett.}\ }\textbf {\bibinfo {volume} {110}},\ \bibinfo {pages}
  {217002} (\bibinfo {year} {2013}{\natexlab{a}})}\BibitemShut {NoStop}%
\bibitem [{\citenamefont {Hinton}\ \emph
  {et~al.}(2013{\natexlab{b}})\citenamefont {Hinton}, \citenamefont {Koralek},
  \citenamefont {Lu}, \citenamefont {Vishwanath}, \citenamefont {Orenstein},
  \citenamefont {Bonn}, \citenamefont {Hardy},\ and\ \citenamefont
  {Liang}}]{Hinton2013}%
  \BibitemOpen
  \bibfield  {author} {\bibinfo {author} {\bibfnamefont {J.~P.}\ \bibnamefont
  {Hinton}}, \bibinfo {author} {\bibfnamefont {J.~D.}\ \bibnamefont {Koralek}},
  \bibinfo {author} {\bibfnamefont {Y.~M.}\ \bibnamefont {Lu}}, \bibinfo
  {author} {\bibfnamefont {a.}~\bibnamefont {Vishwanath}}, \bibinfo {author}
  {\bibfnamefont {J.}~\bibnamefont {Orenstein}}, \bibinfo {author}
  {\bibfnamefont {D.~a.}\ \bibnamefont {Bonn}}, \bibinfo {author}
  {\bibfnamefont {W.~N.}\ \bibnamefont {Hardy}}, \ and\ \bibinfo {author}
  {\bibfnamefont {R.}~\bibnamefont {Liang}},\ }\href {\doibase
  10.1103/PhysRevB.88.060508} {\bibfield  {journal} {\bibinfo  {journal} {Phys.
  Rev. B}\ }\textbf {\bibinfo {volume} {88}},\ \bibinfo {pages} {60508}
  (\bibinfo {year} {2013}{\natexlab{b}})},\ \Eprint
  {http://arxiv.org/abs/1305.1361} {1305.1361} \BibitemShut {NoStop}%
\bibitem [{\citenamefont {Torchinsky}\ \emph {et~al.}(2013)\citenamefont
  {Torchinsky}, \citenamefont {Mahmood}, \citenamefont {Bollinger},
  \citenamefont {Bo\v{z}ovi\'{c}},\ and\ \citenamefont
  {Gedik}}]{Torchinsky2013}%
  \BibitemOpen
  \bibfield  {author} {\bibinfo {author} {\bibfnamefont {D.~H.}\ \bibnamefont
  {Torchinsky}}, \bibinfo {author} {\bibfnamefont {F.}~\bibnamefont {Mahmood}},
  \bibinfo {author} {\bibfnamefont {A.~T.}\ \bibnamefont {Bollinger}}, \bibinfo
  {author} {\bibfnamefont {I.}~\bibnamefont {Bo\v{z}ovi\'{c}}}, \ and\ \bibinfo
  {author} {\bibfnamefont {N.}~\bibnamefont {Gedik}},\ }\href {\doibase
  10.1038/nmat3571} {\bibfield  {journal} {\bibinfo  {journal} {Nat. Mater.}\
  }\textbf {\bibinfo {volume} {12}},\ \bibinfo {pages} {387} (\bibinfo {year}
  {2013})}\BibitemShut {NoStop}%
\bibitem [{\citenamefont {Littlewood}\ and\ \citenamefont
  {Varma}(1981)}]{Littlewood1981}%
  \BibitemOpen
  \bibfield  {author} {\bibinfo {author} {\bibfnamefont {P.~B.}\ \bibnamefont
  {Littlewood}}\ and\ \bibinfo {author} {\bibfnamefont {C.~M.}\ \bibnamefont
  {Varma}},\ }\href {\doibase 10.1103/PhysRevLett.47.811} {\bibfield  {journal}
  {\bibinfo  {journal} {Phys. Rev. Lett.}\ }\textbf {\bibinfo {volume} {47}},\
  \bibinfo {pages} {811} (\bibinfo {year} {1981})}\BibitemShut {NoStop}%
\bibitem [{\citenamefont {Metlitski}\ and\ \citenamefont
  {Sachdev}(2010)}]{Metlitski2010}%
  \BibitemOpen
  \bibfield  {author} {\bibinfo {author} {\bibfnamefont {M.~A.}\ \bibnamefont
  {Metlitski}}\ and\ \bibinfo {author} {\bibfnamefont {S.}~\bibnamefont
  {Sachdev}},\ }\href {\doibase 10.1103/PhysRevB.82.075128} {\bibfield
  {journal} {\bibinfo  {journal} {Phys. Rev. B}\ }\textbf {\bibinfo {volume}
  {82}},\ \bibinfo {pages} {075128} (\bibinfo {year} {2010})}\BibitemShut
  {NoStop}%
\bibitem [{\citenamefont {Sachdev}\ and\ \citenamefont {{La
  Placa}}(2013)}]{Sachdev2013}%
  \BibitemOpen
  \bibfield  {author} {\bibinfo {author} {\bibfnamefont {S.}~\bibnamefont
  {Sachdev}}\ and\ \bibinfo {author} {\bibfnamefont {R.}~\bibnamefont {{La
  Placa}}},\ }\href {\doibase 10.1103/PhysRevLett.111.027202} {\bibfield
  {journal} {\bibinfo  {journal} {Phys. Rev. Lett.}\ }\textbf {\bibinfo
  {volume} {111}},\ \bibinfo {pages} {027202} (\bibinfo {year}
  {2013})}\BibitemShut {NoStop}%
\bibitem [{\citenamefont {Efetov}\ \emph {et~al.}(2013)\citenamefont {Efetov},
  \citenamefont {Meier},\ and\ \citenamefont {P\'{e}pin}}]{Efetov2013}%
  \BibitemOpen
  \bibfield  {author} {\bibinfo {author} {\bibfnamefont {K.~B.}\ \bibnamefont
  {Efetov}}, \bibinfo {author} {\bibfnamefont {H.}~\bibnamefont {Meier}}, \
  and\ \bibinfo {author} {\bibfnamefont {C.}~\bibnamefont {P\'{e}pin}},\ }\href
  {\doibase 10.1038/nphys2641} {\bibfield  {journal} {\bibinfo  {journal} {Nat.
  Phys.}\ }\textbf {\bibinfo {volume} {9}},\ \bibinfo {pages} {442} (\bibinfo
  {year} {2013})}\BibitemShut {NoStop}%
\bibitem [{\citenamefont {Sau}\ and\ \citenamefont {Sachdev}(2014)}]{Sau2013}%
  \BibitemOpen
  \bibfield  {author} {\bibinfo {author} {\bibfnamefont {J.~D.}\ \bibnamefont
  {Sau}}\ and\ \bibinfo {author} {\bibfnamefont {S.}~\bibnamefont {Sachdev}},\
  }\href {\doibase 10.1103/PhysRevB.89.075129} {\bibfield  {journal} {\bibinfo
  {journal} {Phys. Rev. B}\ }\textbf {\bibinfo {volume} {89}},\ \bibinfo
  {pages} {075129} (\bibinfo {year} {2014})}\BibitemShut {NoStop}%
\bibitem [{\citenamefont {Allais}\ \emph
  {et~al.}(2014{\natexlab{a}})\citenamefont {Allais}, \citenamefont {Bauer},\
  and\ \citenamefont {Sachdev}}]{Allais2014}%
  \BibitemOpen
  \bibfield  {author} {\bibinfo {author} {\bibfnamefont {A.}~\bibnamefont
  {Allais}}, \bibinfo {author} {\bibfnamefont {J.}~\bibnamefont {Bauer}}, \
  and\ \bibinfo {author} {\bibfnamefont {S.}~\bibnamefont {Sachdev}},\ }\href
  {\doibase 10.1103/PhysRevB.90.155114} {\bibfield  {journal} {\bibinfo
  {journal} {Phys. Rev. B}\ }\textbf {\bibinfo {volume} {90}},\ \bibinfo
  {pages} {155114} (\bibinfo {year} {2014}{\natexlab{a}})}\BibitemShut
  {NoStop}%
\bibitem [{\citenamefont {Allais}\ \emph
  {et~al.}(2014{\natexlab{b}})\citenamefont {Allais}, \citenamefont {Bauer},\
  and\ \citenamefont {Sachdev}}]{Allais2014a}%
  \BibitemOpen
  \bibfield  {author} {\bibinfo {author} {\bibfnamefont {A.}~\bibnamefont
  {Allais}}, \bibinfo {author} {\bibfnamefont {J.}~\bibnamefont {Bauer}}, \
  and\ \bibinfo {author} {\bibfnamefont {S.}~\bibnamefont {Sachdev}},\ }\href
  {\doibase 10.1007/s12648-014-0488-4} {\bibfield  {journal} {\bibinfo
  {journal} {Indian J. Phys.}\ }\textbf {\bibinfo {volume} {88}},\ \bibinfo
  {pages} {905} (\bibinfo {year} {2014}{\natexlab{b}})}\BibitemShut {NoStop}%
\bibitem [{\citenamefont {Thomson}\ and\ \citenamefont
  {Sachdev}(2015)}]{Thomson2014}%
  \BibitemOpen
  \bibfield  {author} {\bibinfo {author} {\bibfnamefont {A.}~\bibnamefont
  {Thomson}}\ and\ \bibinfo {author} {\bibfnamefont {S.}~\bibnamefont
  {Sachdev}},\ }\href {\doibase 10.1103/PhysRevB.91.115142} {\bibfield
  {journal} {\bibinfo  {journal} {Phys. Rev. B}\ }\textbf {\bibinfo {volume}
  {91}},\ \bibinfo {pages} {115142} (\bibinfo {year} {2015})}\BibitemShut
  {NoStop}%
\bibitem [{\citenamefont {Raines}\ \emph {et~al.}(2015)\citenamefont {Raines},
  \citenamefont {Stanev},\ and\ \citenamefont {Galitski}}]{Raines2015a}%
  \BibitemOpen
  \bibfield  {author} {\bibinfo {author} {\bibfnamefont {Z.~M.}\ \bibnamefont
  {Raines}}, \bibinfo {author} {\bibfnamefont {V.}~\bibnamefont {Stanev}}, \
  and\ \bibinfo {author} {\bibfnamefont {V.~M.}\ \bibnamefont {Galitski}},\
  }\href {\doibase 10.1103/PhysRevB.91.184506} {\bibfield  {journal} {\bibinfo
  {journal} {Phys. Rev. B}\ }\textbf {\bibinfo {volume} {91}},\ \bibinfo
  {pages} {184506} (\bibinfo {year} {2015})}\BibitemShut {NoStop}%
\bibitem [{\citenamefont {Moor}\ \emph {et~al.}(2014)\citenamefont {Moor},
  \citenamefont {Volkov}, \citenamefont {Volkov},\ and\ \citenamefont
  {Efetov}}]{Moor2014}%
  \BibitemOpen
  \bibfield  {author} {\bibinfo {author} {\bibfnamefont {A.}~\bibnamefont
  {Moor}}, \bibinfo {author} {\bibfnamefont {P.~A.}\ \bibnamefont {Volkov}},
  \bibinfo {author} {\bibfnamefont {A.~F.}\ \bibnamefont {Volkov}}, \ and\
  \bibinfo {author} {\bibfnamefont {K.~B.}\ \bibnamefont {Efetov}},\ }\href
  {\doibase 10.1103/PhysRevB.90.024511} {\bibfield  {journal} {\bibinfo
  {journal} {Phys. Rev. B}\ }\textbf {\bibinfo {volume} {90}},\ \bibinfo
  {pages} {024511} (\bibinfo {year} {2014})}\BibitemShut {NoStop}%
\bibitem [{\citenamefont {Chowdhury}\ and\ \citenamefont
  {Sachdev}(2014)}]{Chowdhury2014}%
  \BibitemOpen
  \bibfield  {author} {\bibinfo {author} {\bibfnamefont {D.}~\bibnamefont
  {Chowdhury}}\ and\ \bibinfo {author} {\bibfnamefont {S.}~\bibnamefont
  {Sachdev}},\ }\href {\doibase 10.1103/PhysRevB.90.134516} {\bibfield
  {journal} {\bibinfo  {journal} {Phys. Rev. B}\ }\textbf {\bibinfo {volume}
  {90}},\ \bibinfo {pages} {134516} (\bibinfo {year} {2014})}\BibitemShut
  {NoStop}%
\bibitem [{\citenamefont {Littlewood}\ and\ \citenamefont
  {Varma}(1982)}]{Littlewood1982}%
  \BibitemOpen
  \bibfield  {author} {\bibinfo {author} {\bibfnamefont {P.}~\bibnamefont
  {Littlewood}}\ and\ \bibinfo {author} {\bibfnamefont {C.}~\bibnamefont
  {Varma}},\ }\href {\doibase 10.1103/PhysRevB.26.4883} {\bibfield  {journal}
  {\bibinfo  {journal} {Phys. Rev. B}\ }\textbf {\bibinfo {volume} {26}},\
  \bibinfo {pages} {4883} (\bibinfo {year} {1982})}\BibitemShut {NoStop}%
\bibitem [{\citenamefont {Browne}\ and\ \citenamefont
  {Levin}(1983)}]{Browne1983}%
  \BibitemOpen
  \bibfield  {author} {\bibinfo {author} {\bibfnamefont {D.}~\bibnamefont
  {Browne}}\ and\ \bibinfo {author} {\bibfnamefont {K.}~\bibnamefont {Levin}},\
  }\href {\doibase 10.1103/PhysRevB.28.4029} {\bibfield  {journal} {\bibinfo
  {journal} {Phys. Rev. B}\ }\textbf {\bibinfo {volume} {28}},\ \bibinfo
  {pages} {4029} (\bibinfo {year} {1983})}\BibitemShut {NoStop}%
\bibitem [{\citenamefont {Lei}\ \emph {et~al.}(1984)\citenamefont {Lei},
  \citenamefont {Ting},\ and\ \citenamefont {Birman}}]{Lei1}%
  \BibitemOpen
  \bibfield  {author} {\bibinfo {author} {\bibfnamefont {X.~L.}\ \bibnamefont
  {Lei}}, \bibinfo {author} {\bibfnamefont {C.~S.}\ \bibnamefont {Ting}}, \
  and\ \bibinfo {author} {\bibfnamefont {J.~L.}\ \bibnamefont {Birman}},\
  }\href {\doibase 10.1103/PhysRevB.30.6387} {\bibfield  {journal} {\bibinfo
  {journal} {Phys. Rev. B}\ }\textbf {\bibinfo {volume} {30}},\ \bibinfo
  {pages} {6387} (\bibinfo {year} {1984})}\BibitemShut {NoStop}%
\bibitem [{\citenamefont {Lei}\ \emph {et~al.}(1985)\citenamefont {Lei},
  \citenamefont {Ting},\ and\ \citenamefont {Birman}}]{Lei2}%
  \BibitemOpen
  \bibfield  {author} {\bibinfo {author} {\bibfnamefont {X.~L.}\ \bibnamefont
  {Lei}}, \bibinfo {author} {\bibfnamefont {C.~S.}\ \bibnamefont {Ting}}, \
  and\ \bibinfo {author} {\bibfnamefont {J.~L.}\ \bibnamefont {Birman}},\
  }\href {\doibase 10.1103/PhysRevB.32.1464} {\bibfield  {journal} {\bibinfo
  {journal} {Phys. Rev. B}\ }\textbf {\bibinfo {volume} {32}},\ \bibinfo
  {pages} {1464} (\bibinfo {year} {1985})}\BibitemShut {NoStop}%
\bibitem [{\citenamefont {T\"utt\"o}\ and\ \citenamefont
  {Zawadowski}(1992)}]{Tutto}%
  \BibitemOpen
  \bibfield  {author} {\bibinfo {author} {\bibfnamefont {I.}~\bibnamefont
  {T\"utt\"o}}\ and\ \bibinfo {author} {\bibfnamefont {A.}~\bibnamefont
  {Zawadowski}},\ }\href {\doibase 10.1103/PhysRevB.45.4842} {\bibfield
  {journal} {\bibinfo  {journal} {Phys. Rev. B}\ }\textbf {\bibinfo {volume}
  {45}},\ \bibinfo {pages} {4842} (\bibinfo {year} {1992})}\BibitemShut
  {NoStop}%
\bibitem [{\citenamefont {Cea}\ and\ \citenamefont {Benfatto}(2014)}]{Cea}%
  \BibitemOpen
  \bibfield  {author} {\bibinfo {author} {\bibfnamefont {T.}~\bibnamefont
  {Cea}}\ and\ \bibinfo {author} {\bibfnamefont {L.}~\bibnamefont {Benfatto}},\
  }\href {\doibase 10.1103/PhysRevB.90.224515} {\bibfield  {journal} {\bibinfo
  {journal} {Phys. Rev. B}\ }\textbf {\bibinfo {volume} {90}},\ \bibinfo
  {pages} {224515} (\bibinfo {year} {2014})}\BibitemShut {NoStop}%
\bibitem [{Note1()}]{Note1}%
  \BibitemOpen
  \bibinfo {note} {The oscillations of the phase of superconductor are usually
  pushed up to plasma frequencies by coupling with the Coulomb interaction. In
  contrast, the phase mode of an incommensurate charge order is theoretically a
  Goldstone mode of the system, but in real materials this degree of freedom is
  usually pinned by disorder.}\BibitemShut {Stop}%
\bibitem [{Note2()}]{Note2}%
  \BibitemOpen
  \bibinfo {note} {$\Delta (\tau )$ and $\phi (\tau )$ may be treated, then, as
  real fields since they may always be brought to lie along the real axis via a
  gauge transformation.}\BibitemShut {Stop}%
\bibitem [{Note3()}]{Note3}%
  \BibitemOpen
  \bibinfo {note} {This is a parametrization in terms of longitudinal and
  transverse modes such as considered in Refs.~\protect \rev@citealpnum
  {Littlewood1982,Browne1983} as opposed to radial and angular modes (c.f. D.
  Pekker and C.M. Varma, Annu. Rev. Condens. Matter Phys. 6, 269
  (2015)).}\BibitemShut {Stop}%
\bibitem [{Note4()}]{Note4}%
  \BibitemOpen
  \bibinfo {note} {A similar framework was recently used in Ref.~\protect
  \rev@citealpnum {Cea}. However, the focus of that work was on the effects of
  the superconducting gap on the charge order, and the off-diagonal terms of
  $\protect \mathaccentV {hat}05EQ^R$ (and thus the mixing) were assumed to be
  small.}\BibitemShut {Stop}%
\bibitem [{\citenamefont {Kulik}\ \emph {et~al.}(1981)\citenamefont {Kulik},
  \citenamefont {Entin-Wohlman},\ and\ \citenamefont {Orbach}}]{Kulik}%
  \BibitemOpen
  \bibfield  {author} {\bibinfo {author} {\bibfnamefont {I.}~\bibnamefont
  {Kulik}}, \bibinfo {author} {\bibfnamefont {O.}~\bibnamefont
  {Entin-Wohlman}}, \ and\ \bibinfo {author} {\bibfnamefont {R.}~\bibnamefont
  {Orbach}},\ }\href {\doibase 10.1007/BF00115617} {\bibfield  {journal}
  {\bibinfo  {journal} {Journal of Low Temperature Physics}\ }\textbf {\bibinfo
  {volume} {43}},\ \bibinfo {pages} {591} (\bibinfo {year} {1981})}\BibitemShut
  {NoStop}%
\bibitem [{\citenamefont {Vojta}\ and\ \citenamefont {Sachdev}(2001)}]{Vojta}%
  \BibitemOpen
  \bibfield  {author} {\bibinfo {author} {\bibfnamefont {M.}~\bibnamefont
  {Vojta}}\ and\ \bibinfo {author} {\bibfnamefont {S.}~\bibnamefont
  {Sachdev}},\ }in\ \href {\doibase 10.1007/3-540-44946-9_27} {\emph {\bibinfo
  {booktitle} {Advances in Solid State Physics}}},\ \bibinfo {series} {Advances
  in Solid State Physics Volume 41}, Vol.~\bibinfo {volume} {41},\ \bibinfo
  {editor} {edited by\ \bibinfo {editor} {\bibfnamefont {B.}~\bibnamefont
  {Kramer}}}\ (\bibinfo  {publisher} {Springer Berlin Heidelberg},\ \bibinfo
  {year} {2001})\ pp.\ \bibinfo {pages} {329--341}\BibitemShut {NoStop}%
\bibitem [{Note5()}]{Note5}%
  \BibitemOpen
  \bibinfo {note} {In general, there is no simple time-dependent extension of
  Ginzburg-Landau theory, precisely due to the presence of damping, which
  introduces non-analytic terms (see, for example, I. J. R. Aitchison, G.
  Metikas, and D. J. Lee, Phys. Rev. B 62, 6638 (2000), and references
  therein). We circumvent this difficulty by considering only the $q=0$
  limit.}\BibitemShut {Stop}%
\bibitem [{Note6()}]{Note6}%
  \BibitemOpen
  \bibinfo {note} {We can also add coupling to the lattice degrees of freedom,
  by including bi-linear terms like $g_{\protect \text {ep}} \phi b$ and
  $g_{\Delta } \Delta b$, where $b$ is a phonon mode, and $g_{\protect \text
  {ep}}$ and $g_{\Delta } $ are coupling constants.~\cite {Schaefer2014}
  However, the effects of these couplings appear modest (see the appendix), so
  we will not include them.}\BibitemShut {Stop}%
\bibitem [{\citenamefont {Abrahams}\ and\ \citenamefont
  {Tsuneto}(1966)}]{Abrahams}%
  \BibitemOpen
  \bibfield  {author} {\bibinfo {author} {\bibfnamefont {E.}~\bibnamefont
  {Abrahams}}\ and\ \bibinfo {author} {\bibfnamefont {T.}~\bibnamefont
  {Tsuneto}},\ }\href {\doibase 10.1103/PhysRev.152.416} {\bibfield  {journal}
  {\bibinfo  {journal} {Phys. Rev.}\ }\textbf {\bibinfo {volume} {152}},\
  \bibinfo {pages} {416} (\bibinfo {year} {1966})}\BibitemShut {NoStop}%
\bibitem [{\citenamefont {Pekker}\ and\ \citenamefont
  {Varma}(2015)}]{PekkerVarma}%
  \BibitemOpen
  \bibfield  {author} {\bibinfo {author} {\bibfnamefont {D.}~\bibnamefont
  {Pekker}}\ and\ \bibinfo {author} {\bibfnamefont {C.}~\bibnamefont {Varma}},\
  }\href {\doibase 10.1146/annurev-conmatphys-031214-014350} {\bibfield
  {journal} {\bibinfo  {journal} {Annual Review of Condensed Matter Physics}\
  }\textbf {\bibinfo {volume} {6}},\ \bibinfo {pages} {269} (\bibinfo {year}
  {2015})}\BibitemShut {NoStop}%
\bibitem [{\citenamefont {Larkin}\ and\ \citenamefont
  {Varlamov}(2005)}]{larkin2005theory}%
  \BibitemOpen
  \bibfield  {author} {\bibinfo {author} {\bibfnamefont {A.}~\bibnamefont
  {Larkin}}\ and\ \bibinfo {author} {\bibfnamefont {A.}~\bibnamefont
  {Varlamov}},\ }\href@noop {} {\emph {\bibinfo {title} {Theory of fluctuations
  in superconductors}}}\ (\bibinfo  {publisher} {Clarendon Press},\ \bibinfo
  {year} {2005})\BibitemShut {NoStop}%
\bibitem [{\citenamefont {Sharapov}\ \emph {et~al.}(2001)\citenamefont
  {Sharapov}, \citenamefont {Beck},\ and\ \citenamefont {Loktev}}]{Sharapov}%
  \BibitemOpen
  \bibfield  {author} {\bibinfo {author} {\bibfnamefont {S.~G.}\ \bibnamefont
  {Sharapov}}, \bibinfo {author} {\bibfnamefont {H.}~\bibnamefont {Beck}}, \
  and\ \bibinfo {author} {\bibfnamefont {V.~M.}\ \bibnamefont {Loktev}},\
  }\href {\doibase 10.1103/PhysRevB.64.134519} {\bibfield  {journal} {\bibinfo
  {journal} {Phys. Rev. B}\ }\textbf {\bibinfo {volume} {64}},\ \bibinfo
  {pages} {134519} (\bibinfo {year} {2001})}\BibitemShut {NoStop}%
\bibitem [{\citenamefont {Fu}\ \emph {et~al.}(2014)\citenamefont {Fu},
  \citenamefont {Hung},\ and\ \citenamefont {Sachdev}}]{Hung2014}%
  \BibitemOpen
  \bibfield  {author} {\bibinfo {author} {\bibfnamefont {W.}~\bibnamefont
  {Fu}}, \bibinfo {author} {\bibfnamefont {L.-Y.~Y.}\ \bibnamefont {Hung}}, \
  and\ \bibinfo {author} {\bibfnamefont {S.}~\bibnamefont {Sachdev}},\ }\href
  {\doibase 10.1103/PhysRevB.90.024506} {\bibfield  {journal} {\bibinfo
  {journal} {Phys. Rev. B}\ }\textbf {\bibinfo {volume} {90}},\ \bibinfo
  {pages} {24506} (\bibinfo {year} {2014})}\BibitemShut {NoStop}%
\bibitem [{\citenamefont {Schaefer}\ \emph {et~al.}(2014)\citenamefont
  {Schaefer}, \citenamefont {Kabanov},\ and\ \citenamefont
  {Demsar}}]{Schaefer2014}%
  \BibitemOpen
  \bibfield  {author} {\bibinfo {author} {\bibfnamefont {H.}~\bibnamefont
  {Schaefer}}, \bibinfo {author} {\bibfnamefont {V.~V.}\ \bibnamefont
  {Kabanov}}, \ and\ \bibinfo {author} {\bibfnamefont {J.}~\bibnamefont
  {Demsar}},\ }\href {\doibase 10.1103/PhysRevB.89.045106} {\bibfield
  {journal} {\bibinfo  {journal} {Phys. Rev. B}\ }\textbf {\bibinfo {volume}
  {89}},\ \bibinfo {pages} {045106} (\bibinfo {year} {2014})}\BibitemShut
  {NoStop}%
\end{thebibliography}%

\end{document}